\documentclass[sn-basic]{sn-jnl}

\usepackage{graphicx}%
\usepackage{multirow}%
\usepackage{amsmath,amssymb,amsfonts,aasmacros}%
\usepackage{amsthm}%
\usepackage{mathrsfs}%
\usepackage[title]{appendix}%
\usepackage{xcolor}%
\usepackage{textcomp}%
\usepackage{manyfoot}%
\usepackage{booktabs}%
\usepackage{algorithm}%
\usepackage{algorithmicx}%
\usepackage{algpseudocode}%
\usepackage{listings}%
\usepackage{tcolorbox}
\usepackage{lipsum}
\usepackage{siunitx}
\usepackage{natbib}
\usepackage{enumitem}   
\bibpunct{(}{)}{;}{a}{}{,} 

\newcommand{\Msun}{{\rm\, M_\odot}}
\newcommand{\Msunyr}{{\rm\, M_\odot \, yr^{-1}}}
\newcommand{\pc}{{\rm\, pc}}
\newcommand{\kpc}{{\rm\, kpc}}
\newcommand{\kms}{{\rm \, km \, s^{-1}}}
\newcommand\micron{\hbox{$\mu$m}}
\newcommand{\degree}{^\circ}
\newcommand{\Myr}{\, \rm Myr}
\newcommand{\Gyr}{\, \rm Gyr}

\theoremstyle{thmstyleone}%
\theoremstyle{thmstyletwo}%

\theoremstyle{thmstylethree}%

\begin{document}

\title[Article Title]{Nuclear Stellar Discs}


\author*[1]{\fnm{Mathias} \sur{Schultheis}}\email{mathias.schultheis@oca.eu}
\equalcont{These authors contributed equally to this work.}

\author[2]{\fnm{Mattia~C.} \sur{Sormani}}\email{mattiacarlo.sormani@uninsubria.it}
\equalcont{These authors contributed equally to this work.}

\author[3]{\fnm{Dimitri~A.} \sur{Gadotti}}\email{dimitri.a.gadotti@durham.ac.uk}
\equalcont{These authors contributed equally to this work.}

\affil*[1]{Université Côte d’Azur, Observatoire de la Côte d’Azur, Laboratoire Lagrange, CNRS, Blvd de l’Observatoire, 06304 Nice, France}
\affil[2]{\orgdiv{Como Lake centre for AstroPhysics (CLAP), DiSAT}, \orgname{Universit{\`a} dell’Insubria}, \orgaddress{\street{via Valleggio 11}, \city{Como}, \postcode{22100}, \country{Italy}}}
\affil[3]{Centre for Extragalactic Astronomy, Department of Physics, Durham University, South Road, Durham DH1 3LE, UK}

\abstract{We review our current understanding of nuclear stellar discs (NSDs), rotating, and flattened stellar structures found in the central regions of both early- and late-type galaxies. We examine their demographics, kinematics, stellar populations, metallicity gradients and star formation histories. We derive scaling relations linking NSDs to properties of their host galaxies, and compare them with analogous relations for nuclear star clusters. The relationship between NSDs and other central galactic components, including nuclear rings, nuclear bars, and nuclear star clusters, is explored. The role of NSDs as tracers of the secular evolution of barred galaxies is highlighted, emphasising how they can be used to constrain properties of galactic bars such as their ages. Special attention is given to the Milky Way’s NSD, which serves as a unique case study thanks to its proximity and the ability to resolve individual stars. The review covers the Milky Way’s NSD structure, kinematics, dynamics and stellar content, addressing ongoing debates about the presence of a nuclear bar and implications of the latter for central gas dynamics. We argue that NSDs form by in-situ star formation, most of them because of bar-driven gas inflow, but possibly in some cases because of external acquisition of gas during a gas-rich merger. The review concludes by outlining open questions, future research directions and the exciting prospects provided by upcoming observational facilities. 
}


\maketitle

\section{Introduction} \label{sec:1}

\begin{figure}[t]
\centering
\includegraphics[width=\textwidth]{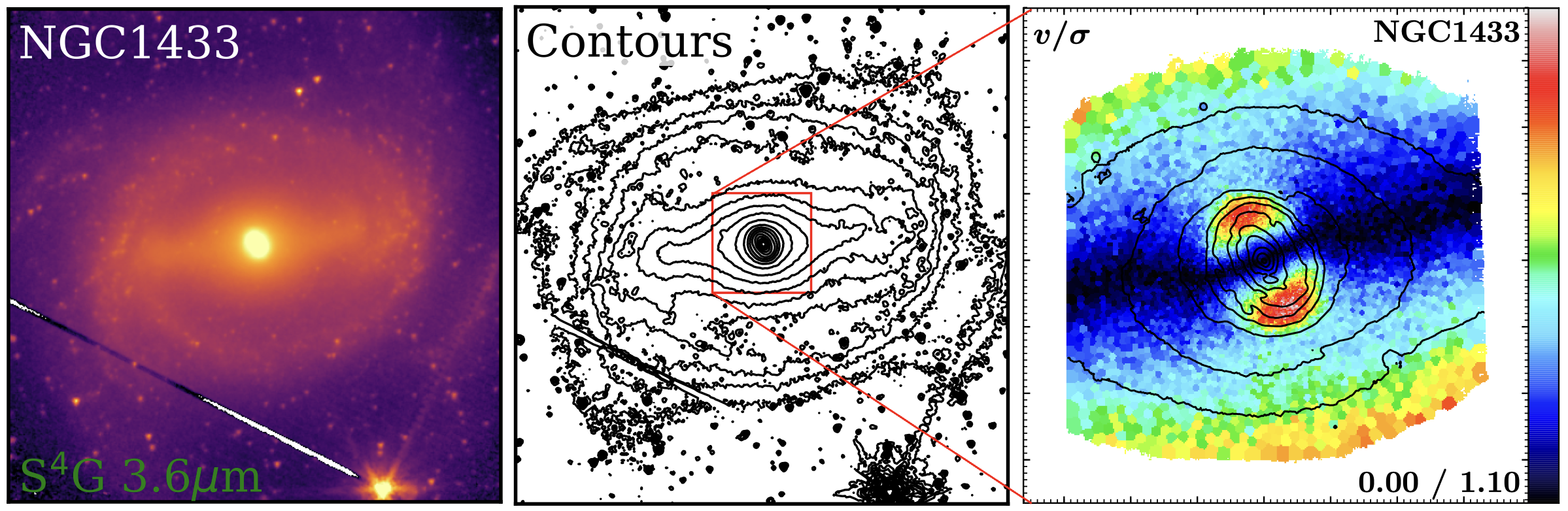}
\caption{Nuclear stellar discs can be identified in external galaxies from photometry (left and middle panels) or from stellar kinematics (right panel). The left panel shows an S$^4$G 3.6$\mu$m image of the barred galaxy NGC 1433. The bar can be seen almost horizontally in this image. The nuclear disc stands out near the centre. The corresponding isophotal contours are shown in the middle panel. The right panel shows a map of stellar $v/\sigma$ derived as part of the MUSE TIMER survey (adapted from \textit{Gadotti, subm.} and \citealt{Gadotti2020}). Stellar kinematics is required to ascertain the unequivocal presence of the nuclear disc as a rapidly-rotating stellar structure.}
\label{fig:NGC1433}
\end{figure}

Nuclear stellar discs (NSDs) are dense, flattened, and rotating stellar systems that can be found in the centre of both early- and late-type galaxies. They have typical radii of a few hundred parsecs \citep[e.g.,][]{Gadotti2020} and typical masses of approximateley a few $10^9\Msun$ \citep[e.g.,][]{deSa-Freitas2025}. Observationally they can be identified photometrically (e.g.,~Fig.~\ref{fig:NGC1433}, left and middle panels) or kinematically (e.g.,~Fig.~\ref{fig:NGC1433}, right panel). They can be nearly axisymmetric, or they can host a non-axisymmetric nuclear bar (e.g.,~Fig.~\ref{fig:Nbars}), similarly to how the main discs in local disc galaxies can host a bar (see Sects.~\ref{sec:nbars} and \ref{sec:nbarformation}). NSDs often contain nuclear star clusters (NSCs) embedded at their centre. 

Although high-resolution images of NSDs have been available since the advent of the \textit{Hubble Space Telescope} (HST; see Sect.~\ref{sec:historical}), interest in them has been accelerating in the last decade mainly driven by the following factors. Firstly, new observational facilities, such as the \textit{Multi-Unit Spectroscopic Explorer} \citep[MUSE;][]{Bacon2010} at the \textit{Very Large Telescope} (VLT), and the \textit{James Webb Space Telescope} (JWST), are allowing to study the morphology and kinematics of dozens of NSDs in detail. Secondly, the realisation that NSDs can give much needed insights on the formation and evolution of bars in disc galaxies (see Sect.~\ref{sec:formation}), and that their presence can affect the evolution of their hosts (including the feeding of Active Galactic Nuclei, or AGN; see Sect.~\ref{sec:nbars}). Thirdly, large-scale infrared surveys of the Galactic centre such as the \emph{GALACTICNUCLEUS} survey \citep{Nogueras-Lara2018} have allowed the study of  the Milky Way's NSD in unprecedented detail, opening a new window into understanding the formation and evolution of our own Galaxy (see Sect.~\ref{sec:MW}). And, finally, the properties of nuclear discs (which we will discuss throughout this paper) suggest that the  central regions of galaxies are significantly shaped by secular evolution, in contrast to the classical picture of bulge formation formed by a random series of disordered merger events (see Sect.~\ref{sec:formation}).

Despite the growing interest, a review of current knowledge on NSDs is lacking. With the advent of future facilities that will provide a wealth of data of unprecedented quality (see Sect.~\ref{sec:perspectives}), there is a need to summarise the state-of-the-art and collect the community's thoughts and discoveries on the subject in a single document, with the goal of fostering further progress. This review is intended to serve both the expert and the newcomer: the former by providing a global perspective, putting together the Milky Way and external galaxies, observation and theory, and the latter by providing a starting point and a list of key open questions that we think will merit attention in the next ten to twenty years. Our summary of the state of the art also attempts at clarifying misconceptions that naturally arise from the development of an incipient field, and today we as a community are in a position to use hindsight to clarify these issues. 
On the way to this goal, we will also try to clarify the issue of confusing nomenclature, which has plagued the study of the central regions of galaxies (Sect.~\ref{sec:nomenclature}).


We organise the review as follows. Section~\ref{sec:historical} gives a short historical summary. Section~\ref{sec:nomenclature} deals with the problem of nomenclature, giving a definition of NSDs and other stellar structures found in the centres of galaxies, in an effort establish a common terminology. Section~\ref{sec:externalgalaxies} gives an overview of NSDs in external galaxies. Section~\ref{sec:MW} reviews our actual understanding of the Milky Way NSD, while in Section~\ref{sec:formation} we discuss the main scenarios for the formation of NSDs. We conclude with a summary of open questions and our view on exciting future opportunities in Section~\ref{sec:perspectives}.

\section{Definitions and Nomenclature} \label{sec:nomenclature}

As attested by the body of work discussed in this review, our understanding of the central regions of disc galaxies has gone through a rapid and multi-faceted evolution in the last twenty years or so. Particularly because new results are coming from different communities, specifically the communities separately working on the Milky Way, nearby galaxies, distant galaxies and theoretical aspects, the lack of standard definitions and of a common nomenclature can -- and have -- become an issue. Use of different language by the different communities to express the same physical concepts or overlapping  physical entities have caused confusion and hampered progress.

For example, the word `bulge' has caused considerable confusion. As described in \citet{Madore2016}, the term became commonly used in astronomy from the 1930s through the 1960s but already in an inconsistent fashion. The essential meaning of the term is a three-dimensional structure that coexists with a comparatively flatter structure: the galaxy bulge `bulges out' of the plane of the main galaxy disc. Taken literally, this definition can only really be directly applied to galaxies seen nearly edge-on. However, perhaps rather unfortunately, a quick association was made between the central structures seen in face-on galaxies with the bulge seen in edge-on galaxies, without a robust justification. A further drawback of the term `bulge' is that it encompasses a large number of different physical structures: for example, the inner part of a bar, a large central spheroidal and even the thick disc, all `bulge out' of the plane of the main galaxy disc.

The term `photometric bulge' is also commonly employed in the context of, e.g., a galaxy intensity (or surface brightness) radial profile. It is associated to any excess of light above the inward extrapolation of the outer disc profile. While useful as a working definition, it is clear that it also encompasses a large number of different physical structures with different formation histories: for example, both a bar and a NSC -- individually or combined -- are `photometric bulges'.

In the following, we give a set of definitions and corresponding nomenclature which will be used throughout this review, with the aim of establishing a common language with which we can effectively describe the state of the field.

\paragraph{Classical Bulge:} The classical picture of the bulge posits that it is a central stellar structure well represented by an oblate, kinematically hot spheroid flattened by rotation (with $v/\sigma<1)$, where $v$ is typically the line-of-sight velocity corrected for inclination, and $\sigma$ is the line-of-sight velocity dispersion\footnote{It is important to note that what matters here are the {\emph{intrinsic} values of $v$ and $\sigma$, which might not always be simple to obtain. Projection effects can be different from intrinsic values  due to the mixing of different components (e.g. bar, NSD) along the line-of-sight.}}. Observationally, they show properties that are to some extent similar to those of elliptical galaxies, with steep surface brightness profiles with a S\'ersic index ${\rm n}\gtrsim2$, pressure supported kinematics, generally rounder isophotes, spheroidal rotation and comprised  primarily by an old, metal-rich stellar population (see \citealt{Barbuy2018} for a review).
Although the early picture included only bulges that stick out of the plane of the disc, more recent work suggests the existence of small classical bulges, fully embedded in the main galaxy disc \citep[see][]{Erwin2015}. One could ask, if they do not bulge out of the disc, why should they be called bulges? Indeed, we agree that they could all be called, e.g., `central hot spheroidal', but at least temporarily it is advantageous to call them `bulges' to connect with the already widely spread concept. 
The formation of classical bulges is not yet fully understood but possibilities include: (i) a relatively rapid collapse of baryons into a dark matter halo (see e.g. \citealt{Eggen1962}, \citealt{Larson1974}, \citealt{Tonini2016}), (ii) hierarchical merging  (see e.g. \citealt{Toomre1977}, \citealt{Cole2000}, \citealt{Querejeta2015}, \citealt{Hopkins2009}, \citealt{Brooks2016}) , and (iii) the central coalescence of disc clumps (see e.g. \citealt{Noguchi1999}, \citealt{Bournaud2007}, \citealt{Perez2013}, \citealt{Devergne2020}, \citealt{Debattista2023}).

\paragraph{Box/Peanut bulge (BP bulge):} In contrast to classical bulges, BP bulges have near-exponential surface brightness profiles, more flattened isophotes, and approximately flat velocity dispersion profiles; they are rotationally supported ($v/\sigma>1$) and have  near-cylindrical rotation \citep{Barbuy2018}. They observationally differ from classical bulges as they have a characteristic boxy shape, sometimes with round vertices resembling a peanut, which occasionally produce a X shape imprinted in images. Terms referring to the same structure include `Box/Peanut', `boxy bulge' and `X-shaped bulge'.
BP bulges are now understood to be the inner part of the bar that grows vertically from within the plane of the main galaxy disc due to dynamical process such as buckling \citep[e.g.,][]{Bureau1999,Bureau2005} or the effects of resonances \citep[e.g.,][]{Combes1990,Patsis2002,Sellwood2020,Baba2022}. Therefore they form after the bar itself. Their orbital structure belongs to the same group of orbital families that constitute the bar. Its typical extent in the radial direction is of the order of one to a few kpc \citep[or about one-third to two-thirds of the full extent of the bar; see][]{Erwin2013}.

\paragraph{Pseudo-bulge:}

This term was originally defined to contrast with classical bulges \citep[see][for a review]{Kormendy2004}. Historically, it has been used to describe any photometric bulge with morphology resembling discs structures, more rotational support, and less concentrated surface brightness profiles than classical bulges \citep[e.g.,][]{Fisher2016}. However, we now understand that this definition includes essentially NSDs and BP bulges \citep[see][]{Athanassoula2005}. The latter are now understood to be the buckled inner parts of bars. We will not adopt this term in this review.

\paragraph{Nuclear disc (ND):} The nuclear disc is a central, kinematically cold structure ($v/\sigma>1$), but typically not as cold as the main galaxy discs (with $v/\sigma\sim5\mhyphen10$). They show a wide range of sizes, from less than 100\,pc to about 1\,kpc \citep{Gadotti2020,deSa-Freitas2023b} and they are clearly a separate structure (i.e., they are not simply the inner part of the main disc). NDs are well described by an exponential intensity radial profile (in a similar fashion as main galaxy discs) and are therefore exponential photometric bulges (i.e., with S\'ersic indices $n\approx1$). For this reason, historically, they have been called `pseudo-bulges', `disc-like bulges' or `discy bulges', to contrast with classical bulges. 
They are thought to be formed from gas brought to the centre for example by a bar or environmental processes (see Sect.~\ref{sec:formation}). This, however, does not exclude the possibility that some NDs are formed via gas accretion in a process unrelated to the bar (see Sect.~\ref{sec:formation}).
We use the term nuclear disc to denote both the stellar and gaseous components.

\paragraph{Nuclear stellar disc (NSD):}

This is the stellar component of a nuclear disc, the focus of this review. Nuclear stellar discs (NSDs) are dense, flattened, and rotating stellar systems that can be found in the centre of both early- and late-type galaxies. They have typical radii of a few hundred parsecs.

\paragraph{Nuclear ring:}

NSDs do not always show ongoing or recent star formation. But when they do, they often show the most intense star formation activity in a ring-shaped region at the edge of the nuclear disc \citep[see][]{Bittner2020}. This is the (\emph{gaseous}) nuclear ring. This is essentially where gas funneled by the bar enters the nuclear disc and is the reason why most star formation is located there, where effects from stellar feedback can also be seen \citep[e.g.,][]{Leaman2019}. The nuclear ring is the outer gaseous component of the nuclear disc. 

\paragraph{Central Molecular Zone (CMZ):}

The term essentially denotes the gaseous part of the nuclear disc. The term originated to designate the accumulation of molecular gas at the centre of the Milky Way, which we now understand to be essentially a star-forming nuclear ring \citep{Morris1996,Henshaw2023}. It is occasionally used to denote similar central accumulation of gas and nuclear rings in external galaxies.



\paragraph{Nuclear Star Cluster (NSC):} NSCs are bright and compact massive stellar clusters with a light excess above the inward extrapolation of the host galaxy's surface brightness profile on scales smaller than about 50 pc \citep{Neumayer2020}. They are some of the densest stellar systems in the universe.

\paragraph{Bar:}

The  bar is a large-scale, elongated stellar structure found in the central regions of many disc galaxies, including the Milky Way \citep[see, e.g.,][who report near-infrared bar fractions of $\sim60$-$75\%$]{Eskridge2000,Menendez-Delmestre2007}. Bars have flat surface density profiles -- which translate to S\'ersic indices $n\leq1$ -- and have semi-major axes in the range from 0.5\,kpc to 10\,kpc, with a typical value of $\approx3\mhyphen5$\,kpc \citep{Erwin2005,Gadotti2011}. Bars are composed of stars in elongated orbits, which introduce a non-axisymmetric component to the overall galaxy potential, with profound consequences. Bars play a crucial role in galaxy dynamics by redistributing angular momentum, channelling gas toward the central regions, and influencing star formation \citep{Athanassoula2013}.

\paragraph{Nuclear bar:}
Nuclear bars are bars found within NSDs, similarly to how large-scale bars are found within galactic discs. 


\paragraph{Nuclear Bulge:}
This is only used for the MW \citep[e.g.][]{Launhardt2002,Nishiyama2013b}, and it means the combination of NSD, NSC, SgrA* and gas in the CMZ.  Since this term is defined primarily by a spatial region in the Milky Way only, and not by physical properties, and because the disc-like nature of some of its major components is contrary to the general concept of a bulge, we will not adopt this term in this review.

\section{Historical perspective} \label{sec:historical}

The modern concept of a NSD as a central, rotating disc-like component distinct from the main galactic disc evolved gradually and morphed at several points in history.
The current notion has become natural only with the availability of high quality data for a larger number of galaxies. In this section we provide a short, and by no means complete, historical account. The history of the Milky Way's NSD proceeded independently and is discussed in Sect.~\ref{sec:MWhistory}.

One could argue that the study of NSDs started as early as the 1970s, for example with the work by \citet{deVaucouleurs1974,deVaucouleurs1975}, who noticed the presence of a nuclear bar in NGC\,1291. This author did not discuss nuclear discs as potential progenitors of nuclear bars, even though numerical studies available at the time indicated that bars form from dynamical instability of discs.

At the beginning of 1980s, it was already understood that bars cause inflow of gas \citep[e.g.][]{Sorensen1976,Prendergast1983} which was later robustly established in numerical hydro-dynamical simulations by \cite{Athanassoula1992}. Based on the measured stellar kinematics of several barred galaxies, \citet{Kormendy1982} proposed that `flat bulges' with disc-like dynamics (which of course are just what we would now call NSDs) can form from gas that has drifted to the central regions during the lifetime of a bar. The picture was still unclear as these structures were also thought to be triaxial. This idea was later further developed into the concept of pseudo-bulges \citep{Kormendy2004}, which we do not adopt in this review (see Sect.~\ref{sec:nomenclature}).

The concept of NSDs was developed further by some theoretical papers. For example, in an effort to explain the feeding of AGN, the seminal paper of \citet[][see also \citealt{Shlosman1990}]{Shlosman1989} put forward the `bars-within-bars' scenario, in which a main bar causes inflow of gas to the central regions of a disc galaxy that builds there a nuclear disc of gas and stars, which then itself develop its own bar, continuing the gas inflow down to the supermassive black hole. As another example of that era, \citet{Pfenniger1990} discussed how dissipation in barred galaxies can cause inflow and the growth of ``central mass concentrations'' (which again are NSDs).

The papers mentioned above are mostly concerned with barred galaxies. In the early 1990s, some papers reported on the detection of NSDs in elliptical galaxies. \citet{Scorza1990} found an embedded disc in an elliptical galaxy using data from the 1.2-m telescope at Calar Alto. In addition, \citet{Cinzano1993} found also evidence for an embedded disc in another elliptical galaxy employing the 3.6-m telescope at La Silla. These and other similar studies were using imaging data and typically finding discy isophotes or an exponential section of the radial surface brightness profile as disc signatures. \citet{Cinzano1993} went a step ahead and decomposed the profile into an embedded disc and a massive spheroid. At this point, the evidence for NSDs in elliptical galaxies was more robust than in disc galaxies, because of their different method of detection. 

At the beginning of the 1990s, the evidence of the existence of NSDs as we understand them today was still sparse and mostly based on data of insufficient quality to draw definite conclusions. The advent of the Hubble Space Telescope (HST) changed this. With much higher spatial resolution than what could be achieved then from the ground, the HST became a crucial tool to study NSDs. Even when the effects of the telescope's early spherical aberration were not yet accounted for, \citet{Jaffe1994} found NSDs in elliptical galaxies with radii of about 100\,pc. In a similar study, \citet{vandenBosch1994} found evidence for NSDs with diameters of about 150\,pc in three lenticular (S0) galaxies. Curiously, they noted that the galaxies were possibly barred, but did not make a connection between the nuclear disc and the bar.

After the HST refurbishment in 1993, with the Wide Field and Planetary Camera 2, \citet{vandenBosch1998a} confirmed the presence of NSDs in more early-type galaxies (elliptical and lenticulars). Taking a significant step further, \citet{Scorza1998} found NSDs in two galaxies with morphological classifications on the border between elliptical and lenticular, and hypothesised that the ``double-disc structure observed in these galaxies has been shaped by now dissolved bars''. This hypothesis was fundamentally based on the suggestion by \citet[][see also \citealt{Emsellem2001}]{Emsellem1996} that an inner disc in the Sombrero galaxy was formed from the infall of pre-enriched gas induced by a bar (see also \citealt{vandenBosch1998b} where this scenario is explored further in the context of another galaxy\footnote{\citet{vandenBosch1998b} also explored the kinematical properties of the galaxy concerned, but their long-slit spectroscopy data was limited.}). \citeauthor{Scorza1998} also asked an important question: are the NSDs just the central part of the main disc, or are the two discs different stellar structures? For the galaxies they studied, they concluded that the latter is true, because their data suggests that the main galaxy disc is almost three times thicker than the observed NSD. Finally, a larger sample of galaxies with HST imaging data was also explored in this context by \citet[][see also \citealt{Morelli2004,Morelli2010,Ferrarese2006,Ledo2010}]{Rest2001}.

While the first studies focused on samples including both elliptical and lenticular galaxies, some work was dedicated to disc galaxies in particular \citep[see, e.g.,][]{Pizzella2002}. This was accompanied by further theoretical work supporting the scenario in which these nuclear rapidly rotating structures were formed through bar-driven processes. In particular, \citet{Michel-Dansac2004} and \citet{Wozniak2007} show simulations in which a kinematically cold and young NSD is formed via bar-driven gas inflow. Finally, another earlier step that shaped the development of the field was given by \citet[][see also \citealt{Balcells2007}]{Balcells2003} who suggested a connection between NSDs and the photometric bulges with low S\'ersic indices in the early bulge/disc decompositions.

\section{Nuclear stellar discs in external galaxies} \label{sec:externalgalaxies}

\subsection{Demographics} \label{sec:frequencies}

Nuclear stellar discs are found in disc galaxies (both spirals and lenticulars) as well as in elliptical galaxies, which do not host a main disc. The most comprehensive statistics available to date on the  distribution of NSDs by galaxy types are obtained from the  Spitzer Survey of Stellar Structures in Galaxies \citep[S$^4$G;][]{Sheth2010}. This is a volume-, magnitude-, and size-limited survey of 2331 galaxies using the Infrared Array Camera (IRAC) at 3.6 and 4.5\,$\mu$m (with distance $d<40$\,Mpc, Galactic latitude $|b|>30^\circ$, apparent B-band magnitude $m_{B\mathrm{corr}}<15.5$, and diameter of the B-band 25\,mag arcsec$^{-2}$ isophote $D_{25}>1$\,arcmin).

Visual morphological classification of all S$^4$G galaxies was performed in detail by \citet{Buta2015}, and includes the classification of the so-called `nuclear variety', which are morphological features commonly associated with NSDs (such as the presence of nuclear bars or nuclear rings and spiral arms). In the following we make the reasonable assumption that the assignment of any such nuclear varieties indicates the presence of an NSD. This allows us to assess how NSDs are distributed as a function of galaxy properties.

\begin{figure}[t]
\centering
\includegraphics[width=\textwidth]{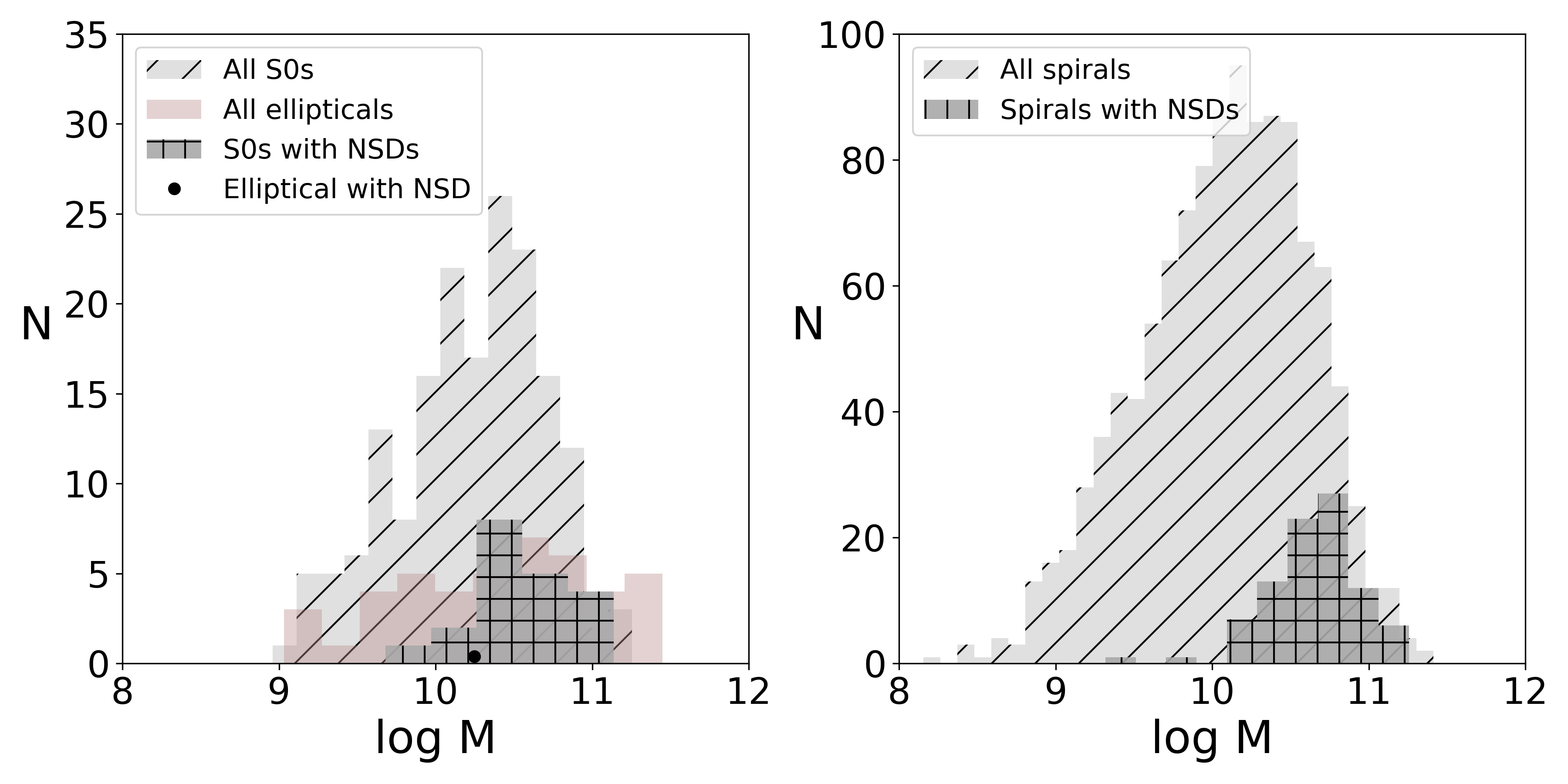}
\includegraphics[width=\textwidth]{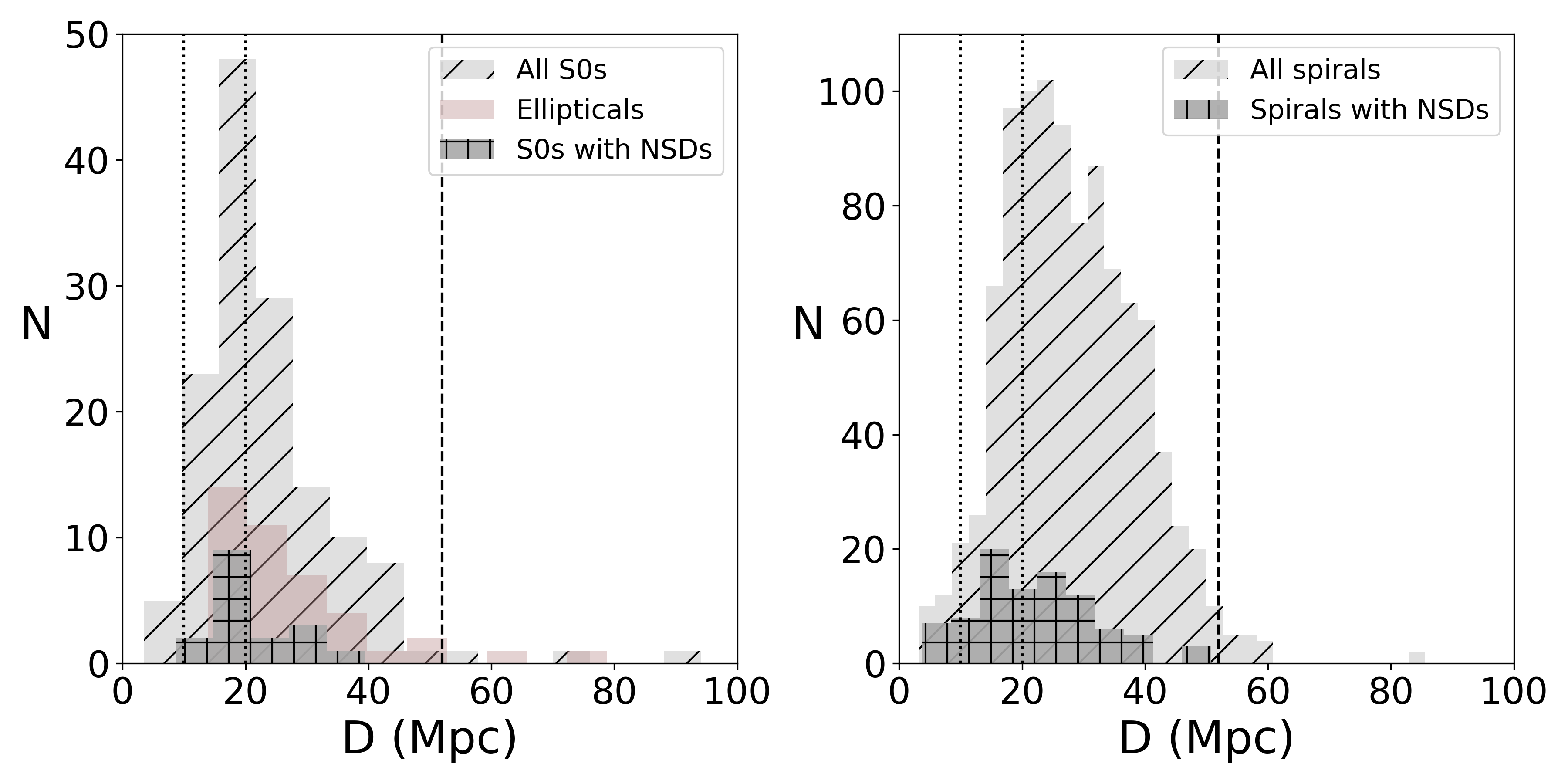}
\caption{\textit{Top row:} distributions of galaxies as a function of total galaxy stellar mass from the S$^4$G survey. Elliptical and lenticular galaxies are on the left, spiral galaxies on the right. \textit{Bottom row:} distribution of galaxies as a function of distance, which may affect our ability to detect NSDs. The dotted and dashed lines mark the distances up to which NSDs with radii of 100, 200 and 500\,pc, respectively, can be detected. Note that the distances employed here are mean redshift-independent distances (from \citealt{Munoz-Mateos2015}), which may be somewhat different from the distances inferred from H{\sc i} radio velocity measurements employed in the original S$^4$G sample selection.}
\label{fig:Mdist}
\end{figure}

The top row in Figure~\ref{fig:Mdist} shows the mass distribution of all galaxies in S$^4$G and the mass distribution of galaxies with NSDs. Taken at face value, the figure suggests that NSDs are found preferentially in more massive systems, for both lenticular and spiral galaxies.  In fact, considering all spiral galaxies, we find that on average $\log(\mathrm{M})=10.07\pm0.02$, whereas when one considers only the spiral galaxies that host NSDs, then, on average $\log(\mathrm{M})=10.65\pm0.03$ (the uncertainties are calculated as the standard error on the mean). The difference is less significant for lenticulars, but it is still there: considering all lenticular galaxies, we find on average that $\log(\mathrm{M})=10.22\pm0.04$, whereas when one considers only the lenticular galaxies with NSDs, then, on average $\log(\mathrm{M})=10.54\pm0.07$. We might ask whether this is a real effect or an observational bias, since our ability to identify small NSDs declines with distance. The bottom panels of Fig.~\ref{fig:Mdist} address this question by showing the distance distribution of galaxies and NSDs in S$^4$G. The vertical dotted and dashed lines mark the distances up to which NSDs with radii of 100, 200 and 500\,pc, respectively, extend over two times the S$^4$G point spread function (PSF) half width at half maximum (HWHM) and can thus be detected. These representative NSDs sizes are chosen because: \textit{(i)} 100\,pc is approximately the current estimate of the radius of the NSD in the Milky Way (see Sect.~\ref{sec:MW}) and the smallest observed NSDs \citep{deSa-Freitas2023b}; \textit{(ii)} 200\,pc is the smallest NSD found in the TIMER sample \citep{Gadotti2019}, and \textit{(iii)} 500\,pc is the mean NSD radius in TIMER. The figure does not show a strong preference for NSDs to be found at shorter distances, which suggests that the prevalence of NSDs in more massive galaxies might be a real phenomenon. However, the figure also shows that while the mean TIMER NSD should be detected with relative ease in S$^4$G, small NSDs can be missed. Thus, we cannot at the moment rule out that NSDs are found preferentially in more massive systems because of observational bias.

\begin{figure}[t]
\centering
\includegraphics[width=\textwidth]{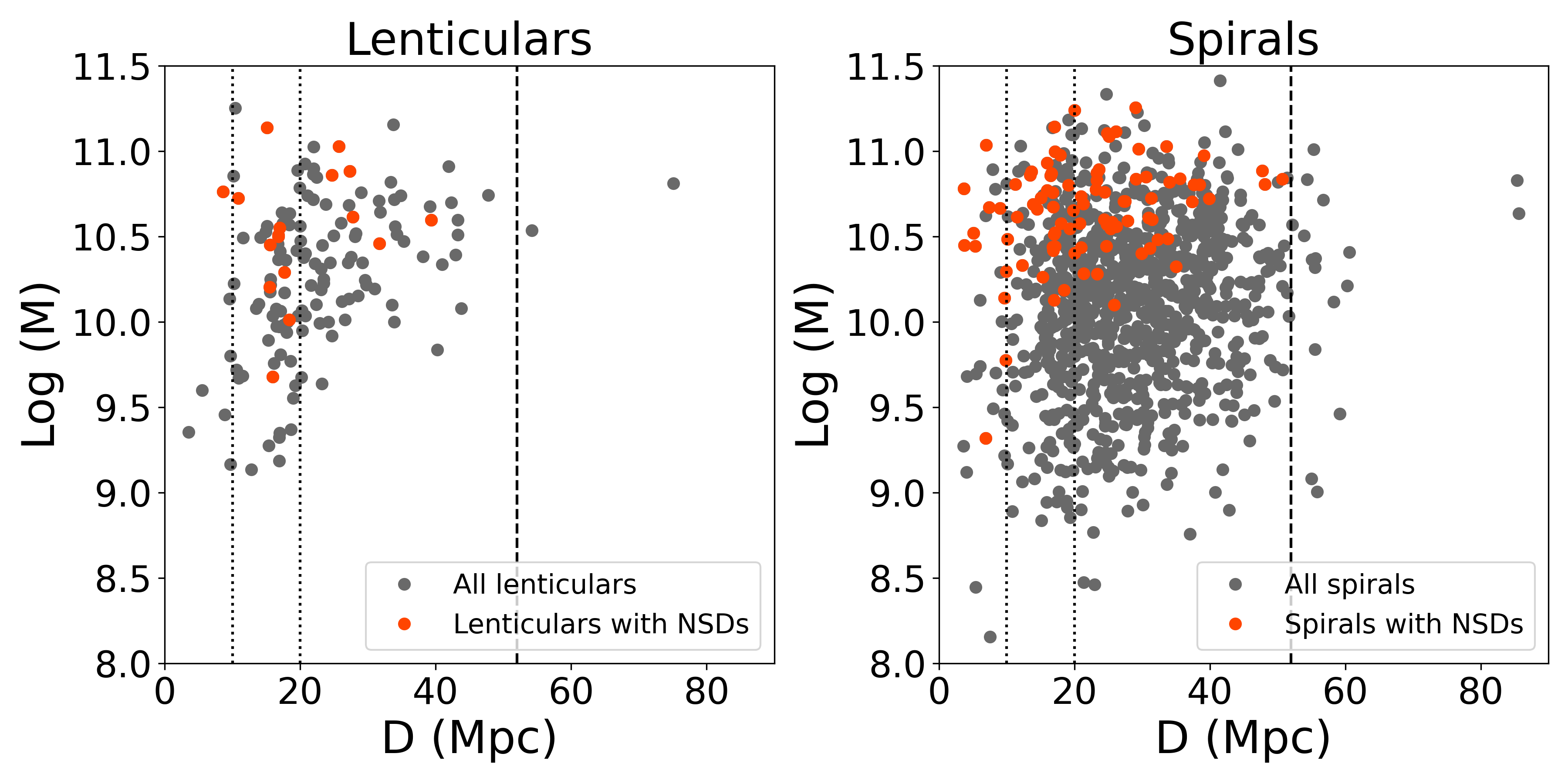}
\caption{ Stellar mass versus distance for lenticular and spiral galaxies in S$^4$G, as indicated. Galaxies hosting NSDs are shown with red filled circles. The vertical lines are the same as in Fig.\,\ref{fig:Mdist}.}
\label{fig:Mvsd}
\end{figure}

\begin{table}[h]
\caption{An assessment of the presence of NSDs in S$^4$G galaxies. The first column specifies the galaxy morphology, while the second column shows the number of galaxies,  the third column shows the number of barred galaxies (and the corresponding fraction of the total number of galaxies with that morphology), the fourth column shows the number of galaxies hosting NSDs (and the corresponding fraction of the total number of galaxies with that morphology), and finally the last column shows the number of galaxies that host both an NSD and a main bar (and the corresponding fraction of the  number of galaxies with that morphology that host NSDs)}. At the bottom of the table, spiral galaxies are further divided into early and late spirals.\label{tab:S4G}%
\begin{tabular}{@{}lcccc@{}}
\toprule
Morphology & Number of galaxies & Bars & NSDs  & NSDs and Bars\\
\midrule
Ellipticals ($\mathrm{T}=-6--4$)     & 58  & 0    & 1  (2\%)  & 0\footnotemark[1]  \\
Lenticulars ($\mathrm{T}=-3.9--1$)   & 197 & 78 (40\%)  & 21 (11\%) & 11 (52\%)  \\
Spirals ($\mathrm{T}=-0.9-7$)        & 1193 & 853 (72\%)   & 90 (8\%)  & 76 (84\%)  \\
\midrule
Early Spirals ($\mathrm{T}=-0.9-3$)  & 426 & 322 (76\%)  & 69 (16\%)  & 61 (88\%)  \\
Late Spirals ($\mathrm{T}=3.1-7$)    & 767 & 531 (69\%)  & 21 (3\%)  & 15 (71\%)  \\
\botrule
\end{tabular}
\footnotetext{A T value of -6 corresponds to the morphological stage cE in \citet{Buta2015}; T=-4 corresponds to E$^+$; T=-3 to S0$^-$; T=-1 to S0$^+$; T=0 to S0/a; T=3 to Sb; and T=7 to Sd. See text for further information on the T parameter.}
\footnotetext[1]{As ellipticals do not host main discs by definition, they also do not host main bars, but an NSD in an elliptical could host a nuclear bar (see Sect.\,\ref{sec:nbars}). We note that the fraction of NSDs in elliptical galaxies may be underestimated in S$^4$G due to spatial resolution effects; see discussion below.}
\end{table}

 In Fig.\,\ref{fig:Mvsd}, we show plots of stellar mass versus distance for lenticular and spiral galaxies in S$^4$G with and without NSDs. One can see again that galaxies hosting a NSD tend to be more massive, although one can also see that not all massive galaxies (regardless of distance) host a visible NSD. Nonetheless, the plots show that below a certain mass ($\approx10^{10.25}\mathrm{M}_\odot$) NSDs are only seen in the more nearby galaxies, which suggests that some NSDs are being missed due to spatial resolution effects. Presumably, NSDs would be systematically smaller in less massive galaxies, implying that these resolution effects would tend to be more prevalent at low masses.

Using S$^4$G, one can also make an assessment of the presence of NSDs in the local universe and how this is related to different morphological classes and the presence of bars. Table\,\ref{tab:S4G} summarises such an assessment, employing the morphological parameter T \citep{Buta2015}, which allows us to quantitatively separate ellipticals, lenticulars (S0s), and early- and late-type spirals\footnote{Table 3 in \citet{Buta2015} provides a detailed view on how the T parameter relates to the various morphological classes.}. The table shows that almost all NSDs are in lenticulars or spiral galaxies, with indeed only one elliptical galaxy showing an NSD  (2\% of the elliptical galaxies in S$^4$G). However, the paucity of NSDs in ellipticals could have been exacerbated if NSDs in elliptical galaxies are significantly smaller than in disc galaxies (i.e., lenticulars and spirals), given the relatively poor S$^4$G spatial resolution of 1.7 arcsec (or approximately 170\,pc at a distance of about 20\,Mpc).  In fact, employing HST photometry, \citet{Ledo2010} find NSDs in 22 galaxies classified as ellipticals (10\% of the elliptical galaxies in their sample), with the vast majority of these NSDs with radii below the spatial resolution provided by the S$^4$G images (while some galaxies are too distant to be included in the S$^4$G sample, as the \citeauthor{Ledo2010} sample goes to 100\,Mpc). Indeed, the average NSD size in the elliptical galaxies in the study of \citeauthor{Ledo2010} is only $\approx110$\,pc. Nonetheless, concerning S0 galaxies, the fraction of NSDs in \citet{Ledo2010} is 19\%, and thus more similar to the fraction in S$^4$G (11\%). The average NSD size in the S0 galaxies in the study of \citeauthor{Ledo2010} is somewhat larger than in the ellipticals, namely $\approx150$\,pc. It is also important to point out that the \citeauthor{Ledo2010} sample is not as statistically significant or complete as the S$^4$G sample, and thus these differences may be due to sample selection effects, different wavelength ranges probed (mid-infrared in S$^4$G vs optical) and NSD detection techniques employed (visual classification in S$^4$G vs ellipse fitting). Therefore, while the S$^4$G analysis is useful to understand the demography of the relatively large NSDs in disc galaxies (S0s and spirals), it should not be taken to reflect a real paucity of NSDs in elliptical galaxies, which appear to be systematically smaller than the NSDs in disc galaxies. This in itself is an interesting observation that could be linked to the different formation processes of NSDs in ellipticals as compared to disc galaxies (see Sect.\,\ref{sec:formation}). Nonetheless, this systematic difference could also result from the various methodological effects in the different samples, and thus needs to be confirmed with further studies.

Table\,\ref{tab:S4G} also shows that most NSDs are in early spirals and lenticulars rather than spirals with morphological stage later than Sb. Further, while the vast majority of spiral galaxies hosting an NSD also have a bar, this fraction drops to about 50\% for lenticulars. This suggests that bars are a primary builder of NSDs and alternative formation scenarios for NSDs could be more efficient in lenticulars.  For example, we speculate that gas accretion from the external environment could reach the inner few 100\,pc or so without the help of a bar (e.g., from an interacting gas-rich companion galaxy), although it is unclear if the interaction would always lead to the formation of bar, or in some cases its weakening \citep[e.g., see discussion in][]{Wang2015,Sauvaget2018}. This could happen more frequently in lenticulars, given that these galaxies tend to live in denser environments as compared to spirals, although this would have to happen early enough so that cold gas is still available to be driven into the galaxy, before other environmental effects preclude this possibility. Alternatively, some studies \citep[e.g.,][]{Diaz-Garcia2016} find that bars are weaker and shorter in lenticulars.  In fact, using the measurements from \citet{Herrera-Endoqui2015}, we find that the mean observed ellipticity of bars in lenticulars is smaller than the corresponding value for spirals, namely, 0.4 and 0.6, respectively. Moreover, these values do not change significantly for galaxies also hosting NSDs. This could lead to bars being missed in the morphological classification, as well as NSDs being missed if they are smaller as a result of the bars being weaker. Therefore, an alternative scenario to explain the lesser association between NSDs and bars in lenticular galaxies would be one in which bars weaken or dissolve more often in lenticulars than in spirals. The bar would still have formed the NSD early on, but would not be readily visible (or not present at all) today. This explanation does not require an alternative path unrelated to bars to form NSDs in lenticulars.  In this context, it is interesting to compare the ratio between the nuclear disc size and the bar size for lenticulars and spirals. We can do that using the measurements from \citet{Erwin2024} for 17 galaxies (seven of which lenticulars), and using the nuclear ring radius as a proxy for the nuclear disc radius (we discuss why this is a valid procedure in Sect.\,\ref{sec:nucstr}). We find that the value of this ratio is remarkably the same, $0.16\pm0.03$, for both lenticulars and spirals that host both a bar and a NSD. This indicates that, at least in the galaxies where a bar is discernible today, the build-up of NSDs in both lenticulars and spirals seems to follow similar bar-driven processes. Inspecting HST images of some of the few unbarred galaxies in S$^4$G showing an NSD, it is clear that while a good fraction of these galaxies appear to be genuine cases, in a number of these galaxies it is unclear whether a bar is present, albeit very weak, or whether the NSD is not simply the inner part of the main disc (i.e., not a stellar structure separate from the main disc). Therefore, the number of genuine cases of unbarred galaxies hosting NSDs is possibly smaller than what is quoted in Table\,\ref{tab:S4G}.

\subsection{Kinematics}
\label{sec:extkin}

NSDs are rotating. They are typically kinematically colder (i.e., more rotationally supported) than hot (pressure supported) spheroids such as classical bulges and BP bulges, but kinematically hotter than main galactic discs. The level of rotational support is usually quantified by the $v/\sigma$ parameter, which is the ratio of the rotation velocity to the velocity dispersion. NSDs typically have $v/\sigma \sim 1\mhyphen3$, while main galactic discs have $v/\sigma \sim 5\mhyphen10$  \citep{Gadotti2020}, although these measurements, particularly for nuclear discs, can be affected by the presence of other structures \citep[but see also][]{deSa-Freitas2023b,Fraser-McKelvie2024}.

Early evidence of these kinematical properties of NSDs include for example \citet{Kormendy1982}, which found that the central regions of several barred galaxies showed amount of rotation sufficient to cause flattening,  and \citet[][see also \citealt{Marquez2003}]{Emsellem2001}, where central `$\sigma$-drops' in the stellar velocity dispersion were found in nearby disc galaxies, using long slit data. This was in contrast to the expectation that the stellar velocity dispersion should rise -- and not decrease -- towards the galaxy centre. 

\citet{deLorenzo-Caceres2008} presented 2D maps of stellar kinematics obtained with the integral field spectrograph (IFS) SAURON for a sample of galaxies with nuclear bars. These maps represented a significant step forward and clearly revealed the kinematically cold nature of NSDs compared to their surroundings. In addition, they showed that at the ends of the nuclear bars one sees substantial drops in the stellar  velocity dispersion. These are not the same $\sigma$-drops discussed above, which are at the centre, and thus these drops were termed `$\sigma$-hollows'.

The stellar kinematic properties of NSDs were more thoroughly revealed -- in terms of sensitivity, field coverage and spatial sampling -- with the Multi-Unit Spectroscopic Explorer \cite[MUSE,][]{Bacon2010} instrument on the Very Large telescope (VLT). \citet[][see also \citealt{Erwin2015}]{Gadotti2015} performed a first pilot study on the lenticular galaxy NGC\,4371 using this instrument, confirming that the NSD is rapidly rotating and has lower velocity dispersion compared to its surroundings. They also measured the $h_3$ and $h_4$ parameters, which are higher order moment of the line of sight velocity distribution (LOSVD) \citep{vanderMarel1993,Gerhard1993}. An anti-correlation between stellar velocity and $h_3$ was clearly seen in the data, indicating that the stellar orbits in the NSD are close to circular. Elevated values of $h_4$ indicated that the NSD sits on a higher velocity dispersion stellar structure, presumably the older, main disc of the galaxy.

More recently, the pilot study of \citet{Gadotti2015} was expanded into the Time Inference with MUSE in Extragalactic Rings (TIMER) survey \citep{Gadotti2019,Gadotti2020}. The TIMER survey with MUSE studied in detail the NSDs in a sample of 21 galaxies. This allowed to further establish the kinematical properties mentioned above and to measure the level of rotational support in NSDs with spatially resolved maps of stellar $v/\sigma$. As mentioned at the beginning of this subsection, the results showed that the inclination-corrected peak $v/\sigma$ in NSDs varies from $\approx1$ to $\approx3$. This contrasts with the level of maximum rotational support in main galaxy discs, which has normal values around $5\mhyphen10$ \citep[see, e.g.,][for the MW disc]{Bland-Hawthorn2016}. This means that, while NSDs are clearly rotationally supported, they can still have a non-negligible pressure support component and have larger height-to-radius ratios and therefore look thicker than main galactic discs.

\subsection{Stellar populations} \label{sec:externalstellarpopulations}

Concerning the gas metallicity of external galaxies hosting NSDs, one of the first spatially-resolved studies was performed by \citet{Sanchez2014} using the  Calar Alto Legacy Integral Field Area Survey (CALIFA), in which they assessed the influence of the bar on the radial gas-phase metallicity gradient. Some of the galaxies in their sample do show an NSD, but their spatial resolution did not allow them to resolve the chemical properties of the gas within the NSDs. \citet{Seidel2016} studied 16 large integral field spectroscopy mosaics of barred galaxies at higher physical spatial resolution and obtained metallicities and [Mg/Fe] abundances from line-strength measurements. They found [Mg/Fe] abundances in bars and central structures higher than in the main disc, suggesting a correlation with the NSD abundances despite their still limited spatial and spectral resolution.

More recently, the TIMER survey  with MUSE at the VLT provided high signal-to-noise ratio (SNR) and relatively high spatial resolution IFS data for the NSDs in 21 nearby barred galaxies \citep{Gadotti2019} that allowed in-depth investigations on the stellar population properties. \citet{Bittner2020} investigated their chemical properties (i.e., metallicities and $[\alpha/{\rm Fe}]$ abundances), concluding that NSDs can be clearly distinguished from their surroundings by their elevated metallicities and depleted $[\alpha/{\rm Fe}]$ abundances. Nevertheless, NSDs are not necessarily composed of stars much younger than those in the surrounding bar or inner main disc. While the results in Bittner et al. show that stellar populations in NSDs have on average younger ages than the immediate surroundings (or contain younger regions, with a typical uncertainty of about a Gyr), the difference may not be very substantial, particularly when the mean stellar ages are measured mass-weighted instead of light-weighted (see, e.g., the case of NGC\,4371, whose NSD has an average stellar age above 10 Gyrs). 

What pushes the measured average stellar age in NSDs to lower values is often the presence of young stars at the outer edge of the NSD. In fact, \citet{Bittner2020} find that NSDs have negative radial gradients of mean stellar age. These gradients are often monotonic, i.e, the mean stellar age decreases from the very centre of the galaxy until the edge of the NSD. At radii larger than the NSD outer edge, one often sees a jump in age, reflecting the more elevated mean stellar ages in the bar and main galaxy disc. Altogether, these properties led Bittner et al. to suggest the inside-out formation scenario for NSDs (see Sect.~\ref{sec:insideout}), in which the central parts form first, and subsequently the disc grows in size with newer stars being formed in a series of gaseous ring of increasing radii, which would naturally lead to the observed gradients. In some cases, the outer edge of the NSD is currently actively forming new stars, which leads to the presence of a star-forming (and often star-bursting) nuclear ring (see following subsection).

The negative gradients in age are typically (but not always) accompanied also by characteristic negative gradients in stellar metallicity and positive gradients of $[\alpha/Fe]$ abundance, both also showing abrupt changes past the NSD \citep{Bittner2020}. These gradients are consistent with the inside-out scenario, indicating a fast chemical enrichment in the central region of the NSD and the evolving population of the outskirts of NSDs, where the onset of Fe-enriching supernovae type Ia is yet to take place more significantly. \citet{Pessa2023} analysed the resolved stellar populations of 19 galaxies from the Physics
at High Angular Resolution in Nearby Galaxies (PHANGS)-MUSE collaboration \citep{Emsellem2022}, with a spatial resolution of $\sim$ 100\,pc. In the galaxies hosting NSDs, these authors confirm the radial gradients of stellar age and metallicity as was found by the TIMER survey.

\subsection{NSDs, nuclear rings and nuclear star clusters}
\label{sec:nucstr}

The presence of star-forming gaseous nuclear rings in disc galaxies has been known for more than half a century, at least since the pioneering work of \citet[][see also \citealt{Buta1996} for an early review on galactic rings]{Sersic1965}. \citet{Comeron2010} provide a more recent and comprehensive study of the photometric properties of nuclear rings in galaxies, including elliptical galaxies. Nuclear rings are often actively forming stars and contain stellar populations that are younger than those in the surrounding regions.

In fact, in some cases the star formation in the nuclear ring is substantial enough to produce effects of stellar feedback on the nearby interstellar medium (ISM), powering galactic winds and outflows \citep[e.g.,][]{Veilleux2020}. For example, in NGC\,3351, \citet{Leaman2019} find warm ionised gas emanating from the star-bursting nuclear ring in NGC\,3351 at approximately 70\,km/s. While the results from the modelling presented in the paper suggest that the gas has insufficient energy to escape the gravitational potential of the galaxy, ALMA observations of the cold phase of the ISM show evidence of shocks between the warm and cold phase. These shocks are shown to affect the kinematics of both cold and warm gas, enhancing their velocity dispersion. with possible negative consequences to the triggering of star formation. The modelling presented in Leaman et al. suggests that direct photon pressure from young stars is more important than supernovae in driving these processes.

Other spectacular examples include NGC\,1365 \citep{Schinnerer2023} and NGC\,1097 \citep{Kolcu2023}. These detailed studies show quantitative evidence for the evolution of star formation in connection to the presence of molecular gas, as well as how the kinematics and ionisation properties of the warm gas are affected by star formation and shocks in these nuclear regions. \citet{Querejeta2021} recently reported high molecular gas and star formation rate surface densities -- comparable to that in spiral arms -- in nuclear rings of the PHANGS sample.

In a recent study of a volume- and mass-limited sample of 155 nearby barred galaxies covering a wide range of Hubble types, \citet{Erwin2024} finds that 20\% of the galaxies in their sample host a nuclear ring. This fraction rises to nearly 50\% for galaxies hosting longer bars. Interestingly, the study reveals that nuclear rings are absent in galaxies with very late Hubble types (Scd-Sd). While there can be observational effects at least partially leading to these results, they can be interpreted in terms of formation and evolution process involving bars, nuclear rings and NSDs (see Sect.\,\ref{sec:formation}).

The importance of nuclear rings in the context of NSDs is that they share the same central location in the inner kpc of disc galaxies. In fact, it was only more recently that the connection between nuclear rings and NSDs has become fully clear. In \citet{Bittner2020}, it was found that nuclear rings are always at the outer edge of NSDs and partake in the monotonic radial gradients of mean stellar age, metallicity and $[\alpha/{\rm Fe}]$ abundance that start at the centre of the galaxy and are a result of the presence of the NSD (see discussion in the previous subsection). This leads to the conclusion that the nuclear rings are naturally and inseparably connected to NSDs; in fact they mark the region of the NSD that first receives the gas inflowing through the bar, which is what leads to the elevated star formation rates often observed.

The connection between NSDs and nuclear star clusters (NSCs) is as yet less clear. At the moment, the observed typical half-light radius of NSDs is approximately in the range 0.2 to 1 kpc (see Fig.\,\ref{fig:ScalRels}; see also, e.g., \citealt{Gadotti2020}), whereas NSCs have half-light radius ranging from a few to almost 100 pc only  \citep[see, e.g.,][]{Boker2002,Cote2006,Georgiev2016,Neumayer2020}. While this difference suggests that these are completely unrelated structures, it is important to point out that observational limitations may be hindering the detection of NSDs with sizes below 100\,pc (see Sect.~\ref{sec:frequencies}). \citet{deSa-Freitas2023b} reports that the smallest NSDs discovered so far (with spectroscopically confirmed rotational support) have outer edges at approximately 100\,pc, suggesting the possibility of an overlap between the largest NSCs and the smallest NSDs.

NSDs and NSCs have different kinematic properties. As mentioned above NSDs have typical $v/\sigma$ in the range $1\mhyphen3$ \citep{Gadotti2020}, while NSCs have $v/\sigma$ typically in the range $0-1$, and are thus more pressure supported, although some with significant rotation \citep[see, e.g.,][]{Seth2008,Seth2010,Feldmeier2014,Nguyen2018,Lyubenova2019,Neumayer2020}.

There are currently two scenarios to explain the formation of NSC. They can form through the merger of star clusters that migrate to the centre due to processes such as dynamical friction \citep{Tremaine1975}; or they can form from in situ star formation at the galactic centre \citep{Milosavljevic2004}, or it can be a combination of both \citep[see, e.g.,][and the review by \citealt{Neumayer2020}]{Hartmann2011,Antonini2012}. \citet{Fahrion2021} find a clear transition in the  relative dominance of these two processes depending on both host galaxy mass and NSC mass. The authors suggest that the migration and merger scenario is responsible for NSCs in dwarf galaxies, whereas central star formation is the leading mechanism that forms NSCs in more massive galaxies. The transition between these two regimes appears to occur at host galaxy masses of approximately $10^9{\rm M}_\odot$, indicating that in most galaxies with evidence for both an NSD and a NSC, the latter was formed predominantly by the in-situ star formation route.

Altogether, these observations naturally lead one to ask: given that bars drive central star formation in disc galaxies, could the formation processes of NSDs and NSCs in massive galaxies be connected? \citet{Nogueras-Lara2023} found some evidence for a smooth gradients of both stellar kinematics and metallicity between the Milky Way's NSD and NSC, and suggested that they might be part of the same structure, part of a joint formation scenario. However, the observations are difficult and need to be confirmed with further studies, also including external galaxies. In addition, it is not ruled out that dynamical mechanisms can lead to such a smooth transition even if the formation processes are unrelated.  We further discuss the relation between NSCs and NSDs in Sect.~\ref{sec:relationNSC}.

\subsection{Scaling relations}
\label{sec:scaling}

\begin{figure}[]
\centering
\includegraphics[width=\textwidth]{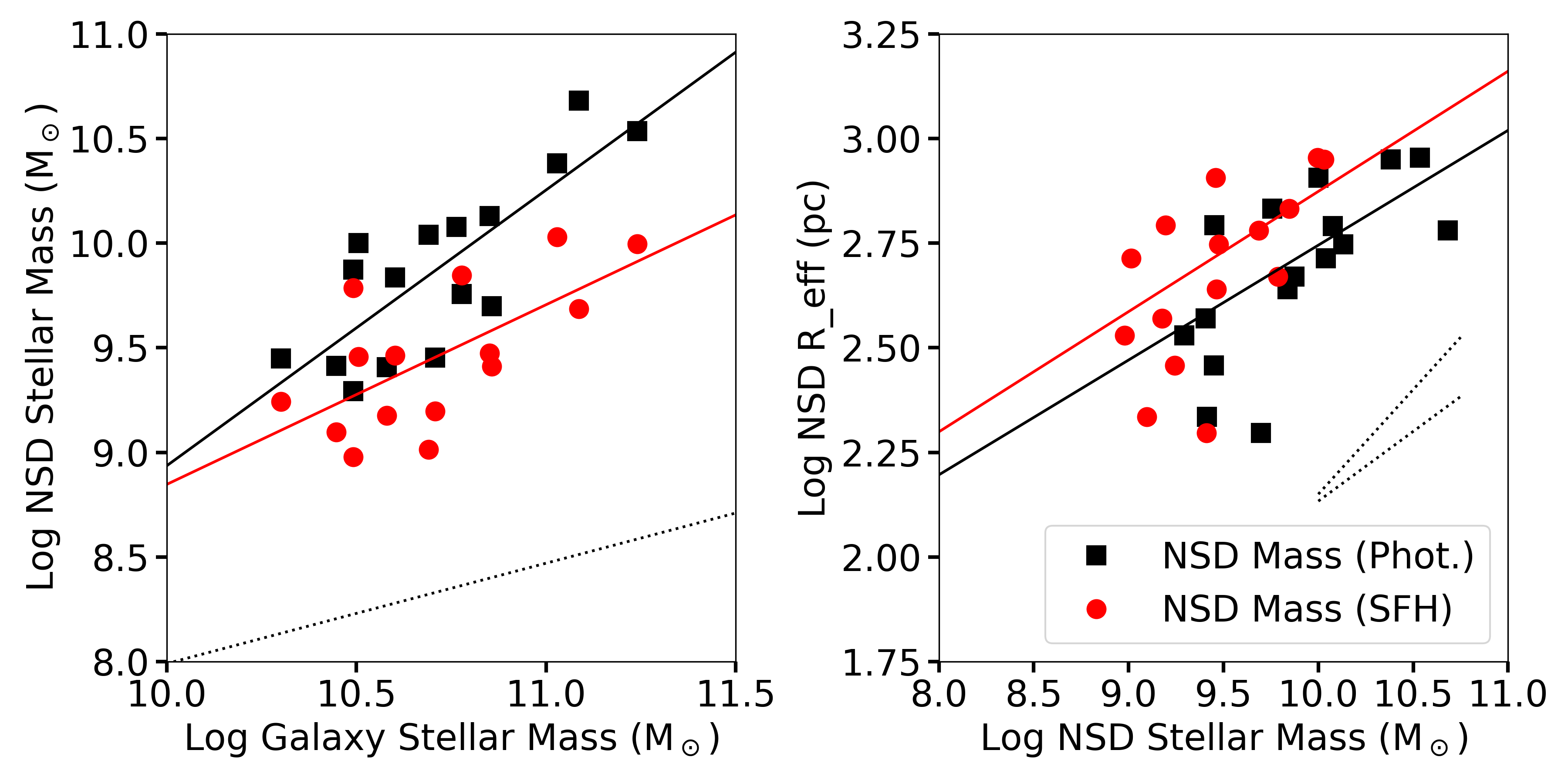}
\caption{\textit{Left:} NSD stellar mass vs.\ host galaxy stellar mass. \textit{Right:} effective (half-light) radius of the NSD vs.\ the NSD stellar mass. Both panels correspond to galaxies in the TIMER sample. NSD radii are derived photometrically from the decompositions in Gadotti (subm.). NSD masses are derived with two different techniques: from the same photometric decompositions (black circles and lines) or via a more sophisticated decomposition of star formation histories \citep[][red circles and lines]{deSa-Freitas2025}. The solid lines in the left panel are fits of the form $y=x^a+b$ and are significantly different from the same relation derived for NSCs (see text for details). The latter is shown as the dotted line in the left panel where the NSC masses are multiplied by 10 to aid visualisation. The solid lines in the right panel are a simple power-law fit, which yields a slope of 0.27 or 0.29 for NSD masses derived photometrically or spectroscopically, respectively. Note that this slope, at face value, is different from the slope for the mass-size relation of NSCs in \citet[][see their Fig.\,7]{Neumayer2020}, which is between 1/3 and 1/2. These two latter slopes are indicated by the dotted segments in the right panel. \textit{(Adapted from Gadotti \& de S\'a-Freitas, subm.)}}
\label{fig:ScalRels}
\end{figure}

Scaling relations are a powerful way to understand formation processes for different stellar structures (see, e.g., \citealt{Graham2009}). For example, one can get insight into the possibility of a joint formation process between NSDs and NSCs by comparing their mass-size relation. \citet{Neumayer2020} found that the mass-size relation of NSCs can be well described by a power law in which the effective (half-light) radius of the cluster is proportional to the cluster mass to a power that is between 1/3 and 1/2. In Fig.\,\ref{fig:ScalRels} we show how the stellar mass of nuclear discs scales with the host galaxy stellar mass (left panel), as well as the NSD mass-size relation (right panel). This figure uses 16 galaxies from the TIMER sample for which a robust photometric decomposition into NSD, bar and main disc was possible using the S$^4$G 3.6$\mu$m images (Gadotti, subm.). These decompositions thus yield the NSD stellar mass as well as its effective (half-light) radius. The galaxy stellar mass is also taken from the S$^4$G derivations \citep{Munoz-Mateos2015}. In addition, the NSD mass can also be obtained via a sophisticated decomposition of the star formation history in the central region of the galaxy, using MUSE spectroscopy. This was done for 15 of these 16 TIMER galaxies in \citet{deSa-Freitas2025} and these results are also shown in the figure. The masses derived spectroscopically are systematically lower than those derived photometrically, but they are strongly correlated, in such a way that the slope of the scaling relations shown here do not depend strongly on which masses are used. Both methodologies have sources of uncertainty stemming, e.g., from contamination from other galactic components, but this may be particularly severe in the case of the photometric estimates, given the limitations in photometric models. This may explain why the photometric estimates are systematically larger than the spectroscopic ones.

The relation between the NSD photometric mass and host galaxy mass is best fitted as:

\begin{equation}
    \log M_{\rm NSD} \approx (1.32\pm0.25) \log \frac{M_{\rm Gal}}{10^9\rm{M}_\odot} + (7.62\pm0.44),
\end{equation}

\noindent while if using the NSD spectroscopic mass this becomes:

\begin{equation}
    \log M_{\rm NSD} \approx (0.86\pm0.25) \log \frac{M_{\rm Gal}}{10^9\rm{M}_\odot} + (7.99\pm0.44),
\end{equation}

\noindent  where the uncertainties are derived via bootstrapping. Note that all galaxies considered in the scaling relations discussed here are massive, disc-dominated galaxies (with stellar masses in the $10 < \log(M) < 11$ range).

These relations are significantly different from the relation found by \citet{Neumayer2020} for NSCs (see their Eq.\,1):

\begin{equation}
    \log M_{\rm NSC} \approx 0.48 \log \frac{M_{\rm Gal}}{10^9\rm{M}_\odot} + 6.51.
\end{equation}

\noindent While this mass scaling relation appears to be very similar for both early- and late-type galaxies, there are indications of a steeper slope when the sample is restricted to more massive late-type galaxies with more accurate NSC mass measurements \citep{Georgiev2016,Neumann2020}. In this case, the slope found is 0.92, which is intermediate between the two measurements for NSDs. At the moment, we lack theoretical predictions that can link formation scenarios to the observed scaling laws. 

While a simple power law is clearly enough to describe the relation between NSD stellar mass and host galaxy stellar mass, the mass-size relation of NSDs appears to be better described by a broken power law. A simple power-law fit yields a slope of 0.27 or 0.29 for NSD masses derived photometrically or spectroscopically, respectively  (with the same uncertainty of 0.003, again derived via bootstrapping). This is close but at face-value incompatible with the slope derived for NSCs by \citet[][i.e., between 1/3 and 1/2]{Neumayer2020}, suggesting that NSDs and NSCs follow formation processes that, although interconnected, are not identical (see below in this paragraph and Sect.~\ref{sec:relationNSC}). In this context, it is important to note that the measurement of NSD masses is prone to large uncertainties, and perhaps the correct values lie between the photometric and spectroscopic results. Nonetheless, at low masses, the NSD mass-size relation appears to follow a steeper relation, although the currently available data is not sufficient to establish this unequivocally. A steeper slope would put the mass-size relation of NSDs closer to that of NSCs. Since low-mass NSDs could be at their initial phases of formation, one could thus speculate and formulate an entertaining possibility; namely, that the formation process of NSDs involve two stages, a first stage in which the NSC and NSD mass grow proportionally to each other, and a second stage in which the NSC and NSD growth is interconnected but in which the NSD grows at a faster rate (see also Sect.~\ref{sec:relationNSC}). Future work should shed light on this possibility.

 Another interesting point to consider is that the nuclear discs shown in Fig.\,\ref{fig:ScalRels} have been formed at different cosmic epochs, and therefore have different ages. In fact, \citet{deSa-Freitas2025} show that these nuclear discs have ages ranging from $\approx1$ to $\approx12$ Gyrs (note that this is not the mean age of the stars in the NSD but the age of the NSD itself). Furthermore, \citet{deSa-Freitas2025} show that the NSD age does not depend on the host galaxy mass (their Fig.\,5), but that it correlates with the NSD size (normalised by the host galaxy size; their Fig.\,7) and the NSD mass normalised by the galaxy mass (their Fig.\,9), which implies that NSDs grow with time. Altogether, this means that the {\em slope} of the scaling relation between the NSD and host galaxy stellar masses (Fig.\,\ref{fig:ScalRels}, left panel) is to a certain degree independent of the NSD age, and that, as NSDs evolve, they move roughly along the scaling relation between their sizes and masses (Fig.\,\ref{fig:ScalRels}, right panel). However, it should be noted that variations in the ages of a sample of NSDs can contribute to the {\em scatter} in the relation between the NSD and host galaxy stellar masses, since older NSDs will tend to populate the upper envelope of the relation. Therefore, biases in the NSD sample selection with respect to NSD age can produce variations in the derived slope.

\subsection{Nuclear bars}
\label{sec:nbars}

\begin{figure}[h]
\centering
\includegraphics[width=\textwidth]{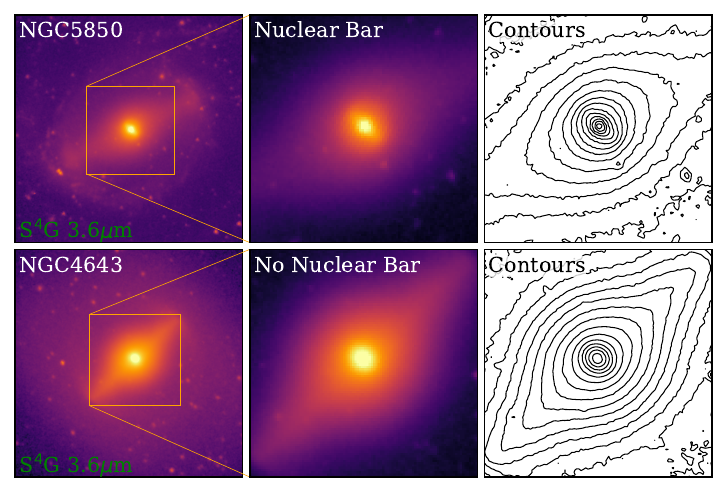}
\caption{NGC\,5850 is an example of a nearby disc galaxy hosting both a main bar and a nuclear bar. The top panels show an S$^4$G 3.6$\mu m$ image (left) where the main bar can be clearly seen as the elongated, diagonal structure. The middle and right panels show, respectively, a cutout of the inner region and the corresponding isophotal contours. The nuclear bar can be clearly seen, almost perpendicular to the main bar. The bottom panels show the corresponding images for NGC\,4643, which, in contrast to NGC\,5850, does not present clear evidence of a nuclear bar in its NSD. \textit{(Adapted from Gadotti, subm.)}}
\label{fig:Nbars}
\end{figure}

The first references to galaxies hosting a second, smaller nuclear bar in addition to the bar hosted by the main galaxy disc are probably from \citet{deVaucouleurs1974,deVaucouleurs1975}, pointing out particularly the case of NGC\,1291. Recent surveys indicate that 20\% of barred galaxies have an additional smaller bar and this fraction rises to about 50\% for galaxies with stellar masses $\log(M_\star/M_\odot)>10.5$ \citep[][see also \citealt{Erwin2004}]{Erwin2024}. Their sizes vary in the range about $100\mhyphen1000$\,pc, very much in the same range of observed NSD radii \citep{Gadotti2020}. Nuclear bars are oriented randomly with respect to main bars \citep{Buta1993,Friedli1993,Erwin2011}, indicating that the pattern speeds of the two bars are likely different. Indeed, direct observational measurements have confirmed that nuclear bars rotate faster than outer bars \citep{Corsini2003,Font2014}. Figure~\ref{fig:Nbars} shows an example of a nearby barred galaxy with an NSD hosting a nuclear bar, as well as a counter example with no clear evidence for a nuclear bar.

As early as the late 1980s, nuclear bars have been suggested as a mechanism to fuel AGN activity \citep{Shlosman1989,Shlosman1990}. While the main bar is able to produce gas inflows from radii about 5\,kpc from the centre to about a few hundred parsecs from the centre \citep[see, e.g.,][]{Sakamoto1999,Sheth2005,daSilva2024}, nuclear bars would continue this process to produce further gas inflows to about 10\,pc from the central supermassive black hole. The literature on the relation between AGN activity and the presence of bars is vast: while many studies find discrepant results \citep[see, e.g.,][]{Mulchaey1997,Ho1997,Knapen2000,Laine2002,Maia2003,Coelho2011}, more recent results suggest that while the presence of a bar is neither a necessary nor a sufficient condition, AGN tend to be preferentially found in barred galaxies \citep{Silva-Lima2022}, or are in galaxies more likely to host a bar \citep{Garland2023}, or tend to be more powerful in barred galaxies \citep{Alonso2013,Alonso2018}. In fact, very recently, \citet{Garland2024} find with robust statistics that the fraction of AGN is higher in barred galaxies. This would suggest that bars play a role at least in building a gas reservoir to fuel AGN, but that further physical processes are required to complete the transfer of gas near enough to the SMBH. Nonetheless, the role of nuclear bars in AGN fuelling is still an open question \citep[e.g.,][]{Martini2001,Maciejewski2002,Laine2002,Erwin2002}. Such studies are complicated by the fact that the time scales related to bar life-times and AGN fuelling and activity are orders of magnitude apart, and the fact that gas availability in the disc may also play a crucial role.

While \citet{deLorenzo-Caceres2020} consider the possibility that nuclear bars may take several Gyrs to form, theoretical work suggest that nuclear bars in NSDs form and evolve just as main bars in main galaxy discs do \citep[e.g.,][]{Maciejewski2007,Maciejewski2008}. In fact, \citet{Mendez-Abreu2019} present observational evidence that the dynamical processes that lead the bar to form a box/peanut bulge also happen in nuclear bars. In addition, \citet{deLorenzo-Caceres2019} find suggestive evidence that nuclear bars can be long-lived, just as main bars can \citep{Gadotti2015,deSa-Freitas2023a,deSa-Freitas2025}, which is consistent with results from both modelling \citep{El-Zant2003} and hydro-dynamical simulations \citep{Wozniak2015}. The longevity of nuclear bars, however, may be cut short if gas inflows are too important, and may not be as large as the longevity of main bars, as suggested by recent simulations \citep{Du2017,Li2023}. The formation of nuclear bars is discussed in more detail in Sect.~\ref{sec:nbarformation}.

As mentioned in Sect.~\ref{sec:extkin}, nuclear bars have been shown to produce drops in the stellar velocity dispersion at their ends known as $\sigma$-hollows \citep[see][]{deLorenzo-Caceres2008}, which can help in identifying such structures. Concerning stellar population properties, \citet{Bittner2021} show that nuclear bars have the same properties as main bars; namely, nuclear bars show elevated values of metallicity ([M/H]) and lower values of alpha-element abundance ([$\alpha$/Fe]), as well lower mean stellar ages at their ends, just as what has been found for main bars by \citet[][see also \citealt{Wozniak2007} for results from simulations concerning mean stellar age]{Neumann2020}.

 The connection between nuclear bars and nuclear discs, and their relation to main bars and main discs, can be explored further by using the measurements in \citet{Erwin2024}. This paper reports on the observed fractions and properties of nuclear bars and nuclear rings (which one can take as the outer edge of a nuclear disc; see Sect.\,\ref{sec:externalstellarpopulations}) in a volume- and mass-limited sample of 155 barred disc galaxies within 30\,Mpc. In this sample, 31 galaxies are identified as hosting a nuclear ring/disc, with 17 (55\%) also hosting a nuclear bar. This fraction is comparable to the fraction of bars observed in main galactic discs (see Sect.\,\ref{sec:nomenclature}). Furthermore, the mean ratio of the nuclear bar size and nuclear stellar disc size in this sample is 0.64, which is somewhat larger (but not too dissimilar) than the corresponding ratio for main bars and main discs: 0.49, as reported in \cite{Gadotti2011}\footnote{It should be noted that, due to spatial resolution limitations, the analysis in \citet{Gadotti2011} misses the shortest bars, i.e., those with a semi-major axis below $2-3$\, kpc.}, and 0.38, as reported in \cite{Erwin2005}. In any case, these numbers should be considered with caution: firstly, some NSDs do not show nuclear rings and are thus not considered in the analysis above; and secondly, limitations due to the spatial resolution and wavelength range of the available data imply that the some nuclear bars have probably been  missed, as a result of being unresolved or difficult to identify due to dust extinction and/or star formation and AGN activity.

Given the abundance of external barred galaxies with NSDs that host a nuclear bar, it would not be surprising if the Milky Way NSD also hosts a nuclear bar. Whether or not it does is an open question and our vantage point in the disc of the Milky Way of course makes it challenging to identify a possible nuclear bar. This is discussed in more detail in Sect.\,\ref{sec:MWnuclearbar}.

\section{The nuclear stellar disc of the Milky Way} \label{sec:MW}

\begin{figure}[h]
\centering
\includegraphics[width=\textwidth]{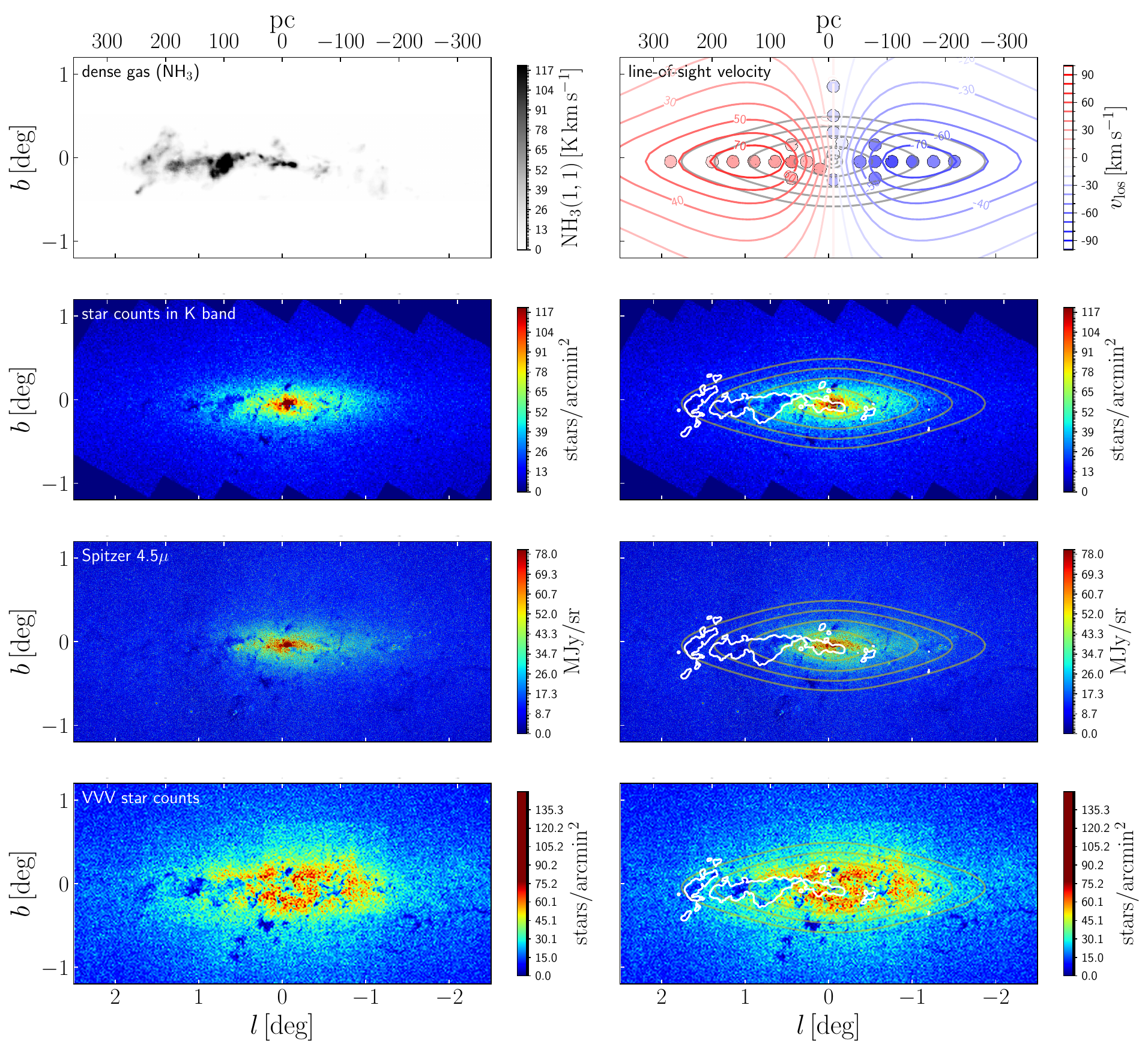}
\caption{\emph{Left column from top to bottom}: 1) Dense gas NH$_3$ (1,1) emission from the HOPS survey \citep{Purcell2012}. 2) Dereddened stellar number density map of the centre of the Milky Way in the K band from \citet{Nishiyama2013b}. 
The NSD is visible as the flattened green structure. The NSC  is the red feature at the centre. 3) Spitzer/IRAC 4.5$\mu$m image \citep{Churchwell2009}. 4) Star counts from the VVV survey using the bump of the AGB ($\rm 11.0 < K_{0} < 12.1$) and a spatial resolution of $\rm 1\,arcmin^{2}$. \emph{Right column from top to bottom}: 1) Circles are the observational fields of the KMOS spectroscopic NSD survey of \citet{Fritz2021}. Circles are coloured by the observed line-of-sight velocity averaged over the field. Coloured contours are the line-of-sight velocity in the best-fit NSD model from \citet{Sormani2022}. Other panels are the same as in the left column with superimposed NH$_3$ emission at 30 $\rm K \, km \, s^{-1}$ (white contours). All panels in the right column show surface density contours of the best-fit model from \citet{Sormani2022}.}
\label{fig:MWnsd}
\end{figure}

It is now well established that embedded at the centre of the Milky Way's bar there is a rotating NSD of total mass $M_{\rm NSD}\simeq 10^9 \Msun$ and exponential radius $R\simeq 90\pc$ (Fig.~\ref{fig:MWnsd}, \citealt{Launhardt2002}, \citealt{Sormani2022}; but see also \citealt{Zoccali2024}). The NSD is flattened with aspect ratio $\simeq 1/3$ and is the dominant component of the Milky Way's gravitational potential in the radial range $30\lesssim R \lesssim 300\pc$. Besides the NSD, the region within Galactocentric radius $R\lesssim 300\pc$ comprises the (much smaller) NSC, the central black hole SgrA* and the ring-like accumulation of gas known as the Central Molecular Zone (CMZ). In fact, the CMZ gas is currently orbiting at radii of $R\simeq 100\mhyphen200\pc$ in the gravitational potential created by the NSD, and star formation occurring in this gas is increasing the NSD mass over time.

The Milky Way's NSD is of special importance because thanks to its proximity it can be studied in much greater detail than any other NSD. It is the only NSD in which we can currently resolve the kinematic and chemical composition of \emph{individual} stars. At the same time, observing the MW's NSD poses special challenges due to our embedded perspective within the Galaxy: the extreme extinction almost completely obscures our view in the optical band ($A_V\simeq 30$), making it necessary to observe in the infrared. The extreme source crowding requires space-based or adaptive optics supported ground-based observations to reach the necessary significantly sub-arcsecond angular resolutions. Because of these technical difficulties, the systematic study of the NSD began only relatively recently (after the year $\sim 2000$), when it became possible to conduct astrometric and spectroscopic infrared observations that cover a significant fraction of the NSD area in the plane of the sky.

In this section, we review the properties of the Milky Way's NSD. We start with a brief historical summary (Sect.~\ref{sec:MWhistory}). We then review its structure \& kinematics (Sect.~\ref{sec:MWkinematics}), including the possibility that the MW's NSD contains a non-axisymmetric nuclear bar (Sect.~\ref{sec:MWnuclearbar}), and finally the current knowledge of its stellar populations and formation history (Sect.~\ref{sec:MWpopulations}).

\subsection{Historical perspective} \label{sec:MWhistory}

\begin{figure}[h]
\centering
\includegraphics[width=\textwidth]{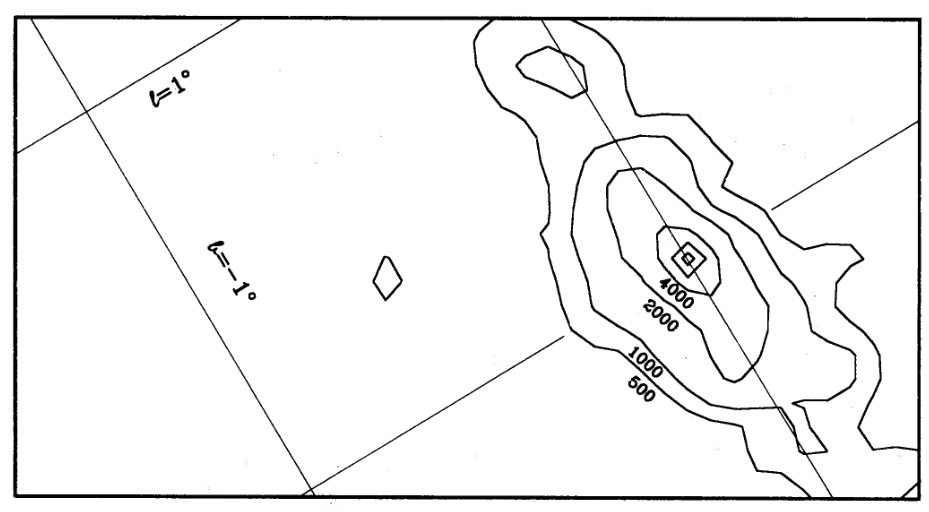}
\caption{The first image of the Milky Way's Nuclear Stellar Disc from \citet{Catchpole1990} (their fig.~7). Numbered contours indicate number of stars per square degree with dereddened $K$-band magnitudes in the range $5<K<6$. The Galactic plane runs diagonally along the line ending at the bottom-right. In hindsight, the flattened structure parallel to the plane is clearly recognisable as the NSD. Compare with more recent star counts in Fig.~\ref{fig:MWnsd}.}
\label{fig:catchpole}
\end{figure}

The realisation that the Milky Way hosts an NSD is relatively recent and happened gradually over time. In hindsight, Fig.~\ref{fig:catchpole}, which reproduces an extinction-corrected star count map of the innermost 2 degrees of the Galactic centre from \citet{Catchpole1990}, can be considered the first image of the MW's NSD. However, at the time it was not obvious that the flattened structure visible in this figure could be identified with an NSD. Indeed, \citet{Catchpole1990} noted that ``the distribution of these stars is clearly elliptical with the major axis along the Galactic plane'', but in the paper did not interpret this flattened distribution as a new disc component. For context, it is useful to remember that the fact that the Milky Way is a barred galaxy became fully established only a year later, from studies of photometry \citep{Blitz1991} and gas dynamics \citep{Binney1991}. 

A couple of years later, \cite{Lindqvist1992} studied the distribution and kinematics of a sample of 134 OH/IR maser stars in the Galactic centre and found that their spatial distribution is similar to that of the star count map of \citet{Catchpole1990} and interpreted the line-of-sight velocities as ``Galactic rotation''. Again that what they were observing was an NSD was not obvious at the time, but in hindsight we can consider this the first kinematic detection of the NSD rotation. We discuss the kinematics of the NSD further in Sect.~\ref{sec:MWkinematics} below.

The first systematic study of the Milky Way's NSD, including the first modern description of its structure, is that of \citet{Launhardt2002}. These authors made a comprehensive analysis of the innermost few degrees of the Milky Way at wavelengths between 2.2 and 240 $\mu$m and identified a ``distinct, massive disk-like complex of stars''  with radius 230 pc, scale height $45$ pc and mass $\sim 10^9 \, \rm M_\odot$. They called this the NSD. This paper established the framework that is still in use today to describe the NSD.

\subsection{Structure and kinematics} \label{sec:MWkinematics}

As mentioned above, the first comprehensive description of the structure of the NSD can be found in \citet{Launhardt2002}. These authors found that the COBE infrared photometry is inconsistent with a single-power law volume density radial profile, but they could obtain a good fit with either a triple-power law profile or a roughly exponential profile. They report a radius of $R=230 \pm 20 \pc$ from COBE data, a vertical scale-height of $H = 45\pm 5\pc$ from warm dust (used as a proxy of the stellar distribution because of the low resolution of the COBE data), and a total mass of $M_{\rm NSD}=1.4 \pm 0.6 \times 10^9 \Msun$. \citet{Nishiyama2013b} fitted an exponential profile to star counts in the $K$ band and found a vertical exponential scale-height of $H \simeq 45 \pm 3 \pc$. \cite{Schodel2014b} found a somewhat smaller scale-height of $H\simeq30\pc$ using Spitzer/IRAC infrared photometry; however, the goal of their work was not to study the structural properties of the NSD but to fit it as a background for their study of the NSC. \cite{Gallego-Cano2020} fitted S{\'e}rsic profiles to the stellar density map from \citet{Nishiyama2013b} and to the Spitzer/IRAC $4.5\micron$ photometry, finding a radial scale-length of $R\simeq90\pc$ and a scale-height of $H\simeq30\pc$. \citet{Sormani2022} constructed axisymmetric self-consistent equilibrium dynamical models of the NSD and fit the models to the 3D kinematics of stars in the NSD KMOS survey of \cite{Fritz2021}. They found roughly exponential radial and vertical scale-lengths of $R = 88.6^{+9.2}_{-6.9} \pc$ and $H=28.4^{+5.5}_{-5.5} \pc$ respectively, and a total mass $M_{\rm NSD} = 10.5^{+1.1}_{-1.0} \times10^8 \,\Msun$. These values are consistent with those of \cite{Launhardt2002} derived purely from photometry (although at first sight the radial and vertical sizes appear to be different, the discrepancy is largely related to the different way in which these scale-lengths are defined, see Sect.~7.2 of \citealt{Sormani2022}), as well as with the other photometry/star-counts estimates listed above, but are completely independent as the fit only used kinematic information and no photometry. Finally, we mention that \cite{Debattista2015,Debattista2018} (see also \citealt{Kunder2021}) argued for the existence of an additional, unrelated and much larger NSD with radius of $\simeq 1\kpc$ as an explanation for the presence of high-velocity peaks in the line-of-sight velocity distribution of stars near the Galactic Centre at $4\degree<l<14\degree$, which however can be also explained by stars on elongated orbits in the Galactic Bar \citep{Molloy2015,Aumer2015,Zhou2021}.

As the NSDs in external galaxies, the MW's NSD is rotating. As mentioned above, the first kinematic detection of the NSD rotation had been accomplished, in hindsight, by \citet{Lindqvist1992} using a sample of 134 luminous OH/IR stars at projected Galactocentric radii $R<100$\,pc. They found  a correlation between the line-of-sight velocity of the stars and the Galactic longitude at $|l|\lesssim 1^\circ$, thus establishing the first evidence of the rotation of the NSD. However, these authors never explicitly mention a ``nuclear disc'' in their paper, generically referring to the detected rotation as ``Galactic rotation''. \citet{Deguchi2004} confirmed their results using a sample of 140 SiO maser stars and attributed the rotation to a ``NSD [that] is rotating more rapidly than the inner bulge''. \citet{Matsunaga2015} measured the line-of-sight velocities of four Cepheid variables in the NSD and found them to be consistent with previously inferred rotation from maser stars. They also noted that the measured line-of-sight velocities are similar to detected gas velocities, suggesting that the cepheids are born in-situ from gas in the CMZ. The rotation of the inner portion of the NSD is also visible in the ISAAC (VLT) near-infrared integral-field spectroscopy map of the central $9.5 \times 8\pc$ of the MW presented in \cite{Feldmeier2014}.

The next major leap in the kinematic study of the NSD came thanks to the APOGEE survey \citep{APOGEE}, which was the first infrared spectroscopy survey of the Galactic mid-plane including the inner Galactic disc, Galactic bar/BP and the Galactic centre. Using this survey \citet{Schoenrich2015} studied the kinematics of M giants situated in the NSD. They selected members of the NSD using a high interstellar extinction cut ($\rm A_{K} > 3$) and found clear evidence of rotation in the NSD  with velocity of $\rm \sim 120\,km\,s^{-1} $. They showed that this rotation is similar to that of the molecular gas (measured e.g. from the CO, $\rm J=4 \xrightarrow{ } 3$ transition, see their Fig.~3) and argued that the NSD kinematics suggests a vertical extent of 50\,pc and a truncation radius of $\rm R \sim 150\,pc$. The similarity of the stellar and gas kinematics corroborates the idea that stars in the NSD are born from gas in the CMZ nuclear ring (See Section~\ref{sec:formation}). \cite{Sormani2020} used a similar APOGEE sample to construct simple axisymmetric Jeans model of the NSD and constrain its mass, finding values consistent with the other studies above.

The largest spectroscopic data set available to date for the NSD has been obtained by \citet{Fritz2021} using the KMOS (ESO/VLT) instrument. Their sample covers more than 3000 stars in the NSD (see their distribution in Fig.~1 of \citealt{Schultheis2021}) for which line-of-sight velocities and stellar parameters such as  effective temperatures and metallicities have been derived (see Sect.~\ref{sec:MWpopulations}). By cross-matching the \cite{Fritz2021} sample with preliminary proper motion data from the VIRAC2 \citep{Smith2025} photometric and astrometric reduction of the Vista Variables in the Via Lactea survey \citep[VVV,][]{Minniti2010}, \citet{Sormani2022} built axisymmetric self-consistent equilibrium dynamical models of the NSD and fitted them to the 3D kinematics (line-of-sight velocities and proper motion). Besides the constraints on mass and density profiles mentioned above, these models also provided an estimate of the relative contribution of bar/NSD stars for each field of the \citet{Fritz2021} KMOS survey, showing that the Galactic bar/BP and NSD contribute $\sim$25\% and $\sim$75\% of all the stars in the innermost fields respectively, and that the bar/BP contribution becomes dominant in the outer fields.

The rotation of the NSD has also been confirmed by \citet{Shazamanian2022} based on a Gaussian decomposition of proper motions parallel to the Galactic plane obtained by combining the GALACTICNUCLEUS data-set and the HST Paschen $\alpha$-survey of the Galactic centre \citep{Dong2011}. \citet{Nieuwmunster2024} investigated in more detail the orbital families that compose the NSD using a probabilistic approach and frequency analysis, finding a large majority of z-tube orbits which are parented $\rm x_{2}$ orbits, in agreement with the studies above and with expectations from the bar-driven scenario of NSD formation (Sect.~\ref{sec:formation}).

The dynamical studies above also demonstrated that the MW's NSD is relatively hot for being a disc (but much colder than the bar/BP), in the sense that the stellar velocity dispersion $\sigma \simeq 70 \kms$ is of the same order as the azimuthally-averaged stellar rotation velocity $\langle v_\phi \rangle\simeq 100\kms$  \citep{Sormani2022,Shazamanian2022}. When the velocity dispersion of stars in a stellar system is non-negligible compared to their rotation velocity, the equilibrium stellar dynamical equations imply that the latter is lower than the circular velocity (i.e., than the velocity of a test particle in a purely circular orbit), an effect called \emph{asymmetric drift} \citep[e.g.][]{Binney2008}. Thus, we expect $\langle v_\phi\rangle$ of stars in the NSD to be $\sim10\mhyphen20\%$ lower than the rotational velocity of the gas in the CMZ (e.g.\, Fig.~18 of \citealt{Sormani2022}). The ratio of the rotational velocity to the intrinsic velocity dispersion of the MW's NSD is therefore of the order $\langle v_\phi\rangle/\sigma \sim 1.4$, in line with the values measured for external NSDs (see Sect.~\ref{sec:extkin}). However, these intrinsic values are difficult to measure due to contamination of the bar.

\subsubsection{Is there a nuclear bar in the Milky Way?} \label{sec:MWnuclearbar}

Whether the MW's NSD contains a nuclear bar (see Sect.~\ref{sec:nbars} for a discussion on nuclear bars in the NSDs of external galaxies) is still under debate and remains an unsolved puzzle. The question is important because, in addition to a revision of our understanding of the NSD structure and dynamics, a nuclear bar would have profound repercussions on our understanding of the gas flows and inflows into the CMZ \citep[e.g.][]{Shlosman1989,Maciejewski2002,Namekata2009,Li2023}. A priori, roughly $40\%$ of external barred galaxies with stellar mass comparable to the MW contain a double bar \citep{Erwin2024}, so the possibility should be taken seriously.

There have been occasional suggestions that the Milky Way is a double barred galaxy. \cite{Alard2001} (see also \citealt{Rodriguez-Fernandez2008}) reconstructed the projected stellar surface density in the plane of the sky from 2MASS star counts and found a positive/negative longitude asymmetry that they interpreted as the signature of a nuclear bar. \cite{Nishiyama2005} and \cite{Gonzalez2011} used red clumps stars as standard candles to trace the bar orientation as a function of longitude, and found a change in slope that they interpreted as the signature of a nuclear bar. However, \cite{Gerhard2012} questioned these results (see also \citealt{Valenti2016})  and showed that both the above signatures could be reproduced by an N-body model that does not contain a nuclear bar.

A major challenge in detecting the presence of a nuclear bar from photometry/star counts is the extreme and non-uniform extinction in the Galactic centre. The distribution of dust and gas in the CMZ is known to be highly asymmetric, with roughly three-quarters of the dust/gas being concentrated at positive longitudes and only one quarter at negative longitudes \citep[e.g.][]{Bally1988,Schultheis2009,Gonzalez2012,Schultheis2014,Nogueras2021,Henshaw2023}. This produces an apparent deficiency of stars at positive longitudes that, if not properly corrected, can be mistakenly interpreted as a nuclear bar oriented with the near side at negative longitudes. Incidentally, this is the same orientation found by \citet{Alard2001}.

An alternative and possibly more reliable way to detect a nuclear bar is through kinematics. Although \citet{Sormani2022} found that the kinematics data in the KMOS NSD survey of \citet{Fritz2021} can be well explained by an axisymmetric model, this could be simply due to low statistics as the survey only contains $\sim 3000$ stars. According to tests performed with N-body models the kinematics signature of a nuclear bar would be rather subtle. In particular, a nuclear bar would look very similar to an axisymmetric NSD in the line-of-sight velocity distribution, but may be detectable with full 3D kinematics information (Fiteni et al. in prep).

A nuclear bar will be harder to detect in the unlucky scenario that its minor/major axis are oriented parallel or perpendicular to the line-of-sight through the Galactic centre (the orientation of a nuclear bar is uncorrelated with that of the main bar, e.g.\, \citealt{Erwin2011}). In this case, no asymmetry whatsoever is expected, although detailed dynamical modelling and/or a boxy-peanut surface density \citep[e.g.][]{Mendez-Abreu2019} might still betray the presence of a nuclear bar.

In summary, current observational constraints appear consistent with an axisymmetric NSD, but the presence of a nuclear bar cannot be ruled out. A larger statistical sample of precise proper motions appears to be the most promising avenue to answer this important question (see Sect.~\ref{sec:perspectives}).

\subsection{Stellar populations} \label{sec:MWpopulations}

\subsubsection{Star formation history} \label{sec:MWsfhistory}

The NSD, with a radius of $R\sim 100\pc$, sits in between the NSC (effective radius $R\sim 5\pc$) and the bar ($R\sim 4\kpc$). Studying the NSD stellar populations is challenging because contamination from either of these two components is almost never negligible. Even in the innermost fields of the \citet{Fritz2021} KMOS survey, contamination from the bar/BP reaches $\sim 20\%$ \citep{Sormani2022}, while at smaller projected radii the contribution of the NSC becomes dominant.

\begin{figure}[h]
\centering
\includegraphics[width=\textwidth]{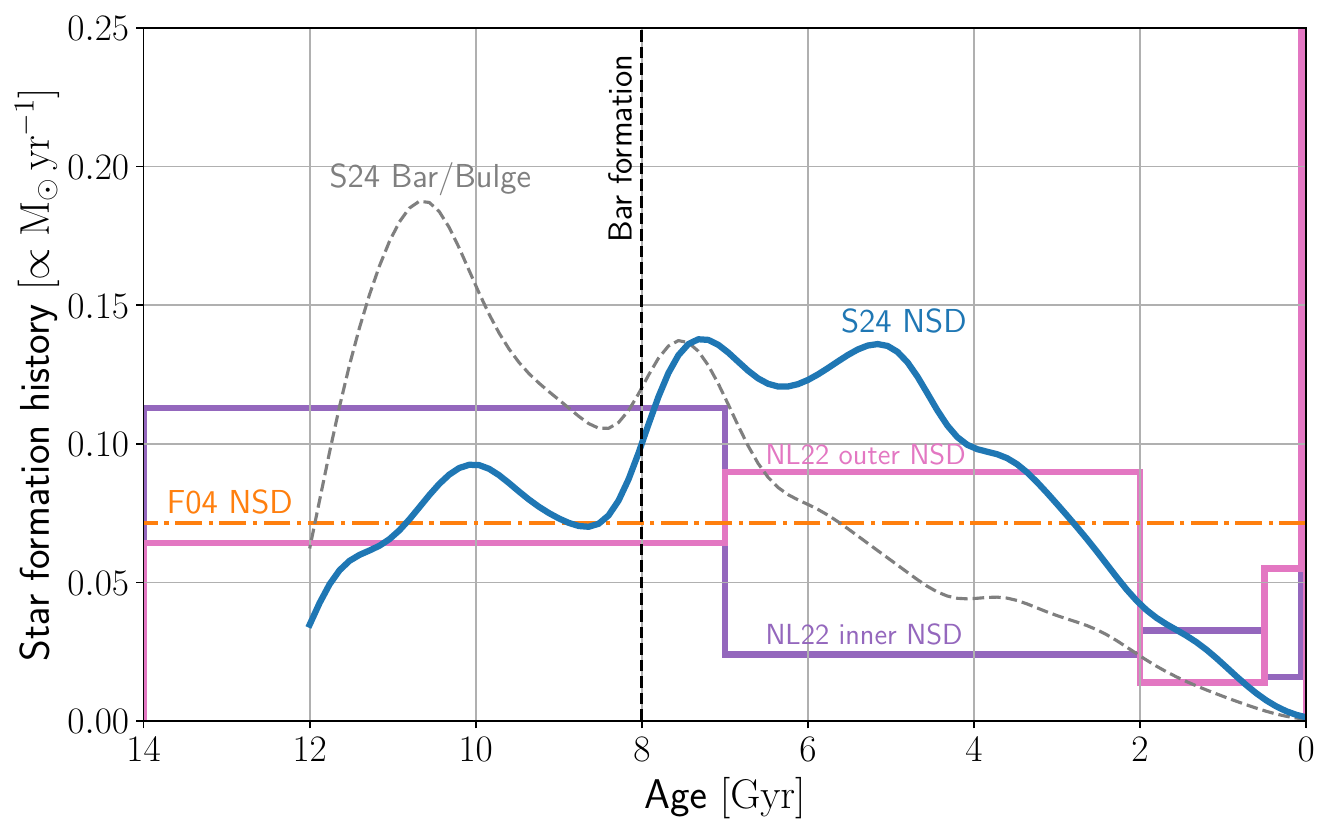}
\caption{Star formation histories (SFHs) of the Milky Way's NSD from the literature. \emph{Blue solid}: from dynamical modelling of Mira variables \citep{Sanders2024}. \emph{Magenta curves:} from modelling the infrared $K$-band luminosity function using stellar evolution models, for the inner and outer part of the NSD separately. The inner NSD curve uses the data from \citet{Nogueras2022SgrB1}. The outer NSD curve is an average of the SgrB1 SFH from \citet{Nogueras2022SgrB1} and of the SgrC SFH from \citet{Nogueras2024}. \emph{Orange dot-dashed}: assuming a constant star formation rate \citep{Figer2004}. \emph{Grey dashed:} SFH of the bar/bulge from dynamical modelling of Mira variables \citep{Sanders2024}. \emph{vertical dashed line:} bar formation epoch according to \citep{Sanders2024}. All curves are normalised so that the total area under the curve is unity.}
\label{fig:SFH}
\end{figure}

Before making a detailed account of the SFH literature, we summarise the current state-of-the-art in Figure \ref{fig:SFH}, which shows recent determinations of the SFH in the NSD. For comparison we also show the SFH for the bar/BP derived from Mira variables in the NSD region from \citet{Sanders2024}, which is in 'reasonable' agreement with the bar/BP SFH of \citet{Bernard2018} from fields in low reddening windows at $\rm -4\degree < b < -2\degree$. The main conclusion of this figure is that there is approximate agreement between the two most recent determinations (see the S24 and NL22 lines) once an average between the inner and outer NL22 results is performed. However, as discussed further below, it is difficult to compare them in detail due to the different temporal resolution and age range of the stellar populations probed.


The first studies of the stellar populations in (what later became understood to be) the NSD are from the late 1980's and 1990's and were mostly concerned with the NSC and/or the bar/BP. These include for example \citet{Rieke1987}, \citet{Lebofsky1987}, \citet{Catchpole1990} and \citet{Blum1995}, who found evidence of an intermediate-age stellar population. \citet{Haller1989} performed a variability study in the central 5'$\times$5' ($12\times12\pc$) that revealed a high number of luminous M giants, possibly young supergiants that imply recent star formation approximately $100 \Myr$ ago.  
\citet{Narayanan1996} sampled the luminosity function on a large area of 16'$\times$16' ($38\times38\pc$) covering a significant fraction of the  NSD and concluded that the region hosts an intermediate-age population. \citet{Sjouwerman1999} probed extreme mass-losing AGB stars (i.e OH/IR maser stars) to find hints of enhanced star formation activity about 1 Gyr ago. This is the first evidence of a non-constant SFH in the NSD, but as the authors cautioned, OH/IR stars are extreme objects with very complicated stellar physics \citep[see e.g.][]{Habing1996}. 

The first dedicated study of the SFH in the Galactic centre on large scales has been carried out by \citet{Figer2004} using deep HST observations across several pencil-beam fields at $|l|\lesssim0.2\degree$ (projected radii $R\lesssim28\pc$) that reach helium-burning red clump stars. They did not distinguish between NSD, NSC and bar/BP. Comparing the observed luminosity functions with stellar evolution models they favoured a constant SFH of $0.02\Msunyr$ in the innermost 30\,pc, and ruled out that the bulk of stars in the region formed 1 Gyr ago in the star burst event hypothesised by \citet{Sjouwerman1999}. 

The next investigation of the SFH in the Galactic centre on large scales, specifically targeting the NSD, was performed by \citet{Nogueras2020NatAs}. They used a method similar to that of \citet{Figer2004} but with the much deeper dataset of the GALACTICNUCLEUS photometric survey. This is the deepest infrared photometric survey so far with a 5\,$\sigma$ detection limit in $\rm J \simeq 22$, $\rm H \simeq 21$ and $\rm K_{S} \simeq 20$ and a  spatial resolution of \SI{0.2}{\arcsecond} \citep{GALACTICNUCLEUS}. \citet{Nogueras2020NatAs} studied a field covering a $90\times30\pc$ field centred on SgrA*, excluding the NSC. They found that the bulk of the NSD  ($\rm \gtrsim 80\%$ of the mass) formed more than 8\,Gyr ago \citep[in agreement with the presence of RR\, Lyrae stars,][]{Minniti2016}, followed by a drop in the star formation activity between $1\mhyphen8$ Gyr ago, and by an increased activity in the last 1 Gyr during which $\sim 5\%$ of the NSD mass formed. They also found a SFR of $\sim 0.1 \Msunyr$ in the last 100 Myr, in accordance with estimates from integrated light measurements (see Sect.~3.1.2 of \citealt{Henshaw2023}), and an elevated SFR of $0.2\mhyphen 0.8 \Msunyr$ in the past 30 Myr, which is several factors higher than the observed present SFR in the CMZ ($\sim 0.1\Msunyr$, Sect.~3.1.1 of \citealt{Henshaw2023}). This large difference in the derived SFR for the young stellar population is caused by large uncertainties and also systematics (due to heavy and patchy interstellar extinction, crowded fields) and have therefore to be considered with caution.

The work of \citet{Nogueras2020NatAs} was expanded and updated by several follow-ups. \cite{Nogueras2022SgrB1} studied the SFH in SgrB1 that at projected radii of $\sim 80\pc$ is located twice as far out as the edge of the region studied in \citet{Nogueras2020NatAs}. \cite{Nogueras2023} studied the SFH history as a function of distance \emph{along} the line of sight using extinction as a proxy for distance. \citet{Schoedel2023} studied the SFH in the Quintuplet cluster field located at a projected radius of $\sim 30\pc$ from SgrA* using HST/WFC3-IR observations. \citet{Nogueras2024} analysed the SgrC region which is located at a roughly symmetrical position of SgrB1 with respect to SgrA*. The key common conclusion from these works is that there is an intermediate-age population (2-7 Gyr old) in the outer (in Galactocentric radius) regions of the NSD that was absent in the inner regions studied by \cite{Nogueras2020NatAs}. This is consistent with the inside-out formation scenario proposed for extragalactic NSDs by \citet{Bittner2020} and discussed more in detail in Sect.~\ref{sec:formation}.

\citet{Sanders2024} determined the SFH of the NSD from dynamical modelling of Mira variables, a class of variable stars with age spanning $1\mhyphen10$~Gyr that can be estimated from a period-age relation (\citealt{Sanders2022}, \citealt{Zhang2023}). Their SFH is shown in Fig.~\ref{fig:SFH} (blue line). Under the assumption that gas was rapidly funnelled to the centre of the Galaxy when the bar formed (e.g.~\citealt{Baba2020} and Sect.~\ref{sec:formation}), they used the derived NSD SFH to deduce that the Galactic bar is $\sim 8 \Gyr$ old (vertical dashed line in Fig.~\ref{fig:SFH}). They found weak or absent evidence for inside-out growth. Their SFH shows approximate agreement with the SFH derived from the GALACTICNUCLEUS series mentioned above (corresponding to NL22 in the figure), once a weighted average between the inner and outer NSD (pink and violet lines) is performed, and once the following uncertainties are taken into account: 1) The elevated SFH found by the GALACTICNUCLEUS papers in the last 1 Gyr cannot be probed with Mira variables as they do not trace reliably ages younger than 1 Gyr; 2) It is unclear to what extent the Period-Luminosity relation calibrated elsewhere in the MW is valid in the NSD; 3) Reliable distance information is not available for the Miras, so there is uncertainty over the location along the line-of-sight.

While the above mentioned studies of the SFH were  mainly focused on either late-type giants and  intermediate-age AGB stars, \citet{Matsunaga2011} and \citet{Matsunaga2015} discovered Cepheid variables in the NSD. From their period-age relation they concluded that these objects are signs of increased star formation during the last 25 Myr. \citet{Yusef-Zadeh2009} studied the entire population of massive YSOs ($\rm 10-20\, M_{\odot}$) in the NSD and derived a present SFR of $\rm 0.14\,M_{odot}$/yr suggesting a recent burst of star formation within the last $\rm 10^{5}$ years.  For a more comprehensive account of measurements of recent ($<100 \Myr$) star formation we refer the reader to the Sect.~3 of the review of \citet{Henshaw2023}.

In summary, the most recent SFH determinations approximately agree, but it is difficult to compare them in detail given that each study only looks at very specific populations (mainly in the later stage of stellar evolution such as M giants or AGB stars). No available study can trace the full stellar populations in the NSD or reach main sequence turnoff stars.
It is evident that a deep spectroscopic survey as well as a deep high angular resolution imaging survey of the NSD is missing to get a global picture of the stellar populations in the NSD. Upcoming instruments such as the MOONS instrument on the VLT \citep{MOONS2011} and the existing NIRCam facility at JWST promise to change this picture in the near future (see Sect.~\ref{sec:upcoming}).

\subsubsection{Metallicities and chemical abundances} \label{sec:abundances}
The gas from which stars are formed is chemically enriched over time by successive generations of stars. The chemical composition of the resulting stellar populations depends on the star formation history, stellar nucleosynthesis, initial mass function  and possible gas flows in/out of the system (\citealt{Matteucci}). Studying the metallicity and chemical composition of stars, which can be measured with stellar spectroscopy, is therefore a powerful tool to trace the formation history of stellar populations.

The KMOS NSD survey of \citet{Fritz2021} provides the largest catalogue of global stellar metallicities in the NSD from mid-resolution spectroscopy ($\rm R\sim 4000$). They derived metallicities for a sample of approximately $\sim$3000 stars using an empirical [Fe/H] calibration using the equivalent widths of CO and Na. Based on this dataset, \citet{Schultheis2021} compared the metallicity distribution of the NSD to that of the NSC \citep{Feldmeier-Krause2020} and the surrounding Galactic bar/BP. Their main conclusions is that the metallicity distribution function (MDF) of the NSD is more metal-rich than the BP but less metal-rich than the NSC, indicating that these three components could be  chemically distinct (see their Fig. 7). Recently, \citet{Feldmeier-Krause2025} presented a global metallicity map including the NSD out to a distance of $\rm \sim 30\,pc$. They see a spatial metallicity variation with a decrease to the center which could be due to a projection effect. By using interstellar reddening as a proxy for distance, \citet{Nogueras-Lara2023} found signs of a metallicity gradient from the NSD down to the NSC, possibly suggesting a smooth transition between these two structures and consistent with expectations from external galaxies and the inside-out formation scenario (see Sect.~\ref{sec:formation}).  Additional evidence of a negative  metallicity gradient comes from a KMOS study of stars in the NSC and the transition region to the NSD.
Linking chemistry and kinematics, \citet{Schultheis2021} and \citet{Nogueras2024a} found that metal-rich stars trace the  kinematically cooler component compared to the metal-poor stars where the velocity dispersion is higher  resulting in a higher fraction of chaotic/box orbits, most likely influences by the Galactic bar.

Regarding detailed chemical abundances, so far there have been a limited amount of measurements in the NSD. The APOGEE spectroscopic survey \citep{APOGEE} could only target the brightest stars of the NSD, such as AGB stars and supergiants for which the chemical abundances are highly uncertain \citep[see e.g.][]{Schultheis2020} due to their complex stellar atmospheres (\citealt{Hoefner2018}). Most of the chemical abundance studies in the Galactic centre mainly focus on the NSC, either using high-resolution spectra \citep[see e.g.][]{Raimrez2000, Rich2017,Thorsbro2020,Bentley2022,Guerco2022, Thorsbro2024, Nishiyama2024} or performing global metallicity studies from low-resolution spectra \citep[i.e. metallicity distribution functions, see e.g.][]{Feldmeier-Krause2017,Feldmeier-Krause2020}. However, as pointed out by \citet{Nieuwmunster2023}, some of the objects studied in these works could belong to the NSD (\citealt{Ryde2016}, \citealt{Nandakumar2018}, \citealt{Thorsbro2020}, \citealt{Rich2017}). A further difficulty is that membership to the NSC, NSD or Galactic bar/BP cannot be attributed on a star-by-star basis, but can only be done in a statistical sense (see e.g. \citealt{Sormani2022}, \citealt{Nogueras2021b}).

The first study of detailed chemical abundance trends was done by \citet{Cunha2007} studying the young supergiant population in the NSC. They found elevated $\alpha$-elements at high metallicities for this population. The majority of subsequent chemical abundances studies focused on the older population of M giants showing alpha-enhancements in the NSC, similar to the thick disc and the Galactic bar/BP (\citealt{Ryde2015}, \citealt{Ryde2016}, \citealt{Nandakumar2018}, \citealt{Nieuwmunster2023}). \citet{Do2018} derived high abundances of V, Sc and Y in two cool metal-rich stars. \citet{Thorsbro2020} found evidence of Si-enhancement for their metal-rich sample of stars in the NSC indicating some possible sign of a starburst, but their sample consists of very cool M giants ($\rm T_{eff} < 4000\,K$) where  due to the presence of numerous molecular bands, the chemical abundance analysis is extremely challenging. Recently, much effort has been achieved to increase the precision of chemical abundances by establishing a differential analysis of stars in the NSD/BP with respect to a well constrained solar neighbourhood sample (see e.g. \citealt{Nieuwmunster2023}, \citealt{Nandakumar2024}), which allows one to pin down the precision of e.g.\ [Si/Fe] to 0.05\,dex. In this spirit, \citet{Ryde2024} analysed chemical abundances of M giants situated in the NSC where they could not confirm the presence of a recent star-burst, but cannot exclude that there might be a complex high-metallicity environment  with a spread in [Si/Fe] values. Obviously larger high precise quality samples (see Sect.~\ref{sec:perspectives}) are necessary to probe the chemical evolution history of the NSD. \citet{Kovtyukh2019} and \citet{Kovtyukh2022} analysed chemical abundances of four cepheid variables in the NSD and found that these have an iron content close to the solar value.

The lack of chemical abundances in the NSD makes it difficult to compare with predictions from chemical evolution models such as those of \citet{Grieco2015} and \citet{Friske2023}. These authors developed the first chemical evolution models of NSDs, with a particular focus on the MW's NSD. Their models are spatially resolved and can make predictions for metallicity and chemical abundances profiles under various formation and accretion scenarios. Obtaining more chemical abundance data such  as $\rm \alpha$-elements, Fe-peak elements, r and s-process elements would help constrain the star formation history and gas flow pattern of the NSD.

\section{Formation and Evolution of Nuclear Discs} \label{sec:formation}

NSDs in all types of galaxies (spirals, lenticulars and ellipticals; see Sect.~\ref{sec:frequencies}) are flat and rotationally supported. Because of this basic property, they are usually assumed to form predominantly by in-situ star-formation, in contrast to NSCs which can be built both by in-situ star formation and by inward migration and merging of globular clusters \citep[see][]{Neumayer2020,Fahrion2021}. Indeed, a large number of globular clusters drawn from an approximately isotropic distribution migrating to the centre would tend to give rise to dynamically hot spheroidal systems with small or moderate amounts of rotation \citep[e.g.][]{Hartmann2011,Antonini2012,Tsatsi2017} and not to flat  strongly rotating dynamically cold discs, although it cannot be excluded that the inward migration of star clusters plays a minor role in the formation of some NSDs as well \citep{Portaluri2013,Arca-Sedda2018}.

In-situ star formation requires gas infall to the centre. The mechanism by which the gas falls into the centre to form an NSD may vary depending on the galaxy. In the case of barred galaxies, which can be either spirals or lenticulars and constitute the majority of galaxies known to have an NSD (see Sect.~\ref{sec:frequencies}), it seems natural to assume that the bar is responsible for the gas infall, since bars are known to efficiently transport gas to the centre. We explore bar-driven NSD formation in more detail in Sect.~\ref{sec:insideout}.

In the case of unbarred galaxies, such as ellipticals, the situation is more uncertain. One possibility is that NSDs in unbarred galaxies \emph{also} form in barred systems, which have then undergone a merger event that destroyed the bar while preserving the integrity of the NSD, creating an elliptical or otherwise unbarred galaxy with an NSD at the centre. In this scenario, whose feasibility still needs to be demonstrated, the NSD should be older than the merger. If such a merger interrupts early the inside-out growth of the NSD in the progenitor barred galaxy (see Sect.~\ref{sec:insideout}), or if the outer regions of an NSD are stripped away during the merger leaving only the central components intact, it might provide an explanation of why NSDs in elliptical galaxies appear to be systematically smaller (see Sect.~\ref{sec:frequencies}). N-body simulations currently disfavour this scenario and suggest that NSDs are frequently fully disrupted during major mergers, although they can more easily survive minor or intermediate mergers \citep{Ledo2010, Sarzi2015, Galan-deAnta2023}. A more comprehensive investigation of the merger parameter space (e.g. orientation, impact parameter, mass ratio) with high-resolution self-consistent N-body models is probably required to fully evaluate the likelihood of NSD survival, and we cannot currently completely rule out the possibility that all NSDs are formed in bars.

Another way to explain the gas infall and formation of NSDs in unbarred galaxies is external acquisition of gas during a gas-rich (minor or major) merger \citep{Pizzella2002,Morelli2004,Mayer2008,Corsini2012}. In this case, the NSD should be younger or have the same age of the merger, and could be used as a clock to determine the age of the merger event \citep{Ledo2010,Sarzi2016,Corsini2016}. An NSD formed in this way should lack any radial gradient in age or metallicity, in contrast to the well-defined radial gradients observed in many barred galaxies (see Sect.~\ref{sec:insideout}). The absence of such radial gradients may be a way to distinguish between NSDs formed by bar-driven infall or gas-rich mergers. Finally, stellar mass loss ($\rm \sim 10^6 M_{\odot}$ of gas) coming from stellar winds or pair-instability supernovae in the region surrounding the NSD could be responsible for some gas inflow towards the centre \citep{Bailey1980,Bailey1982}.

\subsection{Bar-driven formation} \label{sec:bardriven}

Galactic bars are common, being present in $\sim 2/3$ of spiral galaxies in the local Universe \citep[e.g.][]{Erwin2018}. Recent observations from JWST have shown that stellar bars are also frequent at higher redshifts \citep[e.g.][]{Guo2023,Costantin2023,Amvrosiadis2024,Tsukui2024,LeConte2024}. As discussed in Sect.~\ref{sec:externalgalaxies}, all barred galaxies appear to possess an NSD when observed at sufficiently high resolution. Because NSDs are so ubiquitous in barred galaxies, and for the further reasons discussed below, it seems very likely that NSDs in these galaxies form as a consequence of bar-driven inflow.

Galactic bars drive radial inflows of gas towards the centre of galaxies. Their gravitational torques efficiently remove angular momentum from the gas and transport it from radii of several kpc down to the central few hundred pc, where they often produce ring-like accumulations of gas. The transport of gas happens primarily through the two so-called bar “dust lanes”, which are effectively kpc-long streams of gas, one on each leading side of the bar \citep[e.g.][]{Athanassoula1992,Tress2020}. In external barred galaxies, the central ring-like accumulation of gas is usually called a nuclear ring \citep[e.g.][]{Comeron2010,Stuber2023}. In the Milky Way, it is better known as the ``Central Molecular Zone" \citep{Morris1996,Henshaw2023}.

Stars in NSDs are most likely born from gas in nuclear rings, as substantiated by various lines of observational evidence. For example, the size of gaseous nuclear rings often coincides with the outer edge of NSDs, and the rotation velocities of gas and stars are very similar, both for external galaxies \citep{Gadotti2020,Bittner2020} and the Milky Way \citep{Schoenrich2015,Schultheis2021,Sormani2022}. Furthermore, NSDs are typically younger, more metal-rich, and show lower $[\alpha/{\rm Fe}]$ enhancements than their immediate surroundings, as expected if NSDs form from the bar-driven gas inflows, both in external galaxies \citep{Bittner2020} and the Milky Way (Sect.~\ref{sec:abundances}).

In the bar-driven formation scenario, the bar inflow rate as a function of time controls the growth of the NSD \citep{Cole2014,Seo2019}. Observations suggest that typically inflow rates in the present day are in the range $0.1$-$10 \, \rm M_\odot\, yr^{-1}$ \citep[e.g.][]{Regan1997,Laine1999,Elmegreen2009,Shimizu2019,Hatchfield2021,Sormani2023} and typical SFR of nuclear rings are in a similar range \citep{Mazzuca2008,Ma2018}. At these rates, an NSD of mass $M_{\rm NSD}\sim 10^{9}\mhyphen 10^{10}\rm M_\odot$ can form on a timescale of $\sim 1\mhyphen 10 \, \rm Gyr$. However, not all the inflowing mass is necessarily converted into star formation as galactic outflows and winds may remove gas from the nuclear ring before it has time to form stars \citep[see, e.g.,][]{Leaman2019}.

\subsubsection{Inside-out evolution} \label{sec:insideout}

The most promising framework for describing the growth of NSDs in barred galaxies is the inside-out evolution scenario proposed by \citet{Bittner2020}. These authors used the MUSE TIMER survey to derive 2D maps of mean stellar ages, metallicities, and $[\alpha/{\rm Fe}]$ abundances in a sample of 21 barred galaxies. They found that many galaxies show well-defined radial gradients, and in particular that ages decrease with radius up to the outer edge of NSDs. This led them to propose that NSDs are formed from a series of star-forming gaseous nuclear rings that grow in radius over Gyr timescales. In this picture, nuclear rings are the growing outer edge of NSDs.

This inside-out formation scenario is well supported by theoretical considerations. To understand why, we need to first understand where the bar-driven inflow deposits the gas. As mentioned above, the bar inflow deposits the gas in a nuclear ring,  where it settles onto nearly-circular orbits. But what controls the radius of the nuclear ring? This turns out to be a non-trivial question in galactic dynamics. In brief, the nuclear ring is an accumulation of gas at the inner edge of a large ``gap'' or ``cavity'' that forms around the inner Lindblad Resonance (ILR) of a rotating bar potential \citep{Sormani2024}. The process is similar to the opening of the Cassini division in Saturn's rings by the Mimas satellite \citep{Goldreich1978} and to the opening of gaps by an embedded planet in protoplanetary rings \citep{Lin1993}. The important point for our discussion is that the radius of the ring correlates with (but is smaller than) the radius of the ILR, and depends on the parameters of the bar potential in a similar way. In particular, the ILR moves out and the radius of the nuclear ring increases if one or more of the following occurs \citep[e.g.][]{Athanassoula1992,Kim2012,Li2015}:
\begin{enumerate}[label=(\roman*)]
\item The total  mass in the  nuclear region (i.e., the mass of the NSD) increases.
\item The pattern speed of the bar decreases.
\end{enumerate}
During the secular evolution of barred galaxies both these processes work together in increasing the size of nuclear rings over time. The total mass in the nuclear region increases as the bar brings in more gas and the NSD grows, moving the average position of the gaseous nuclear ring outwards over Gyr timescales. 
The bar pattern speed decreases over secular timescales due to dynamical friction with the dark matter halo \citep[e.g.][]{Weinberg1985,Athanassoula2003,Chiba2023}, also resulting in inside-out growth. Note that the radius of a nuclear ring increases during the secular evolution of a galaxy not because the gas of the ring physically expands, but because the fresh gas brought in by the bar forms a new nuclear ring with a radius slightly larger than the previous one. These effects and the inside-out growth of a simulated NSD are beautifully illustrated in the simulations of \citet{Seo2019}, see in particular their Fig.~15.

We note that an NSD could also form from the dissolution of a nuclear bar (see \citealt{Wozniak2015} and Sect.~\ref{sec:nbarformation}). In this case, a nuclear bar formed in the early stages might dissolve into an axisymmetric NSD that then continues to grow inside-out.

While the most convincing observational evidence in support of the inside-out scenario is for external galaxies, some tentative evidence exists also for the Milky Way. \citet{Nogueras-Lara2023} used extinction as a proxy for distance to find evidence that the outer parts of the NSD have a more significant intermediate-age populations of 2–7 Gyr than the inner parts, which supports an inside-out growth. \citet{Sanders2024} analysed Mira variables in the NSD in different age groups and found some weak evidence that the radius of the NSD increases with age. The situation in the MW is made more challenging than for external galaxies by our embedded perspective within the Galactic disc. More observations of ages, metallicities and chemical abundances in both the MW and nearby galaxies will be necessary to further confirm or disprove the inside-out theory.

\subsubsection{Nuclear stellar discs as tracers of the evolution history of galactic bars} \label{sec:evohistory}

As mentioned in the introduction, one of the reasons for the growing interest in NSDs in the last decade is the realisation that they could give insights on the formation and evolutionary history of galactic bars and their host galaxies. 

NSDs can help constrain the age of the bar. \citet{Gadotti2015} first proposed that the oldest stars in an NSD should be slightly younger than the bar in a bar-driven NSD formation scenario, and used this to estimate that the bar in NGC 4371 was already in place at $z\sim 1.8$. The TIMER survey was then built on this principle \citep{Gadotti2019}, i.e., to use the star formation history of the NSD to derive the age of the bar in external barred galaxies, leading \citet{deSa-Freitas2023a} to develop the first generally applicable observational methodology that employs the star formation histories of both the NSD and the main underlying disc of the galaxy to estimate bar ages, and applying this in a pilot study to find that the bar of NGC 1433 is approximately 7.5 Gyr old. From a theoretical perspective, numerical simulations predict that the formation of the bar triggers a boost in the bar-driven inflow rate and therefore in the star formation rate for a period typically lasting around 1 Gyr, during which a significant fraction of the total NSD mass can be built \citep{Cole2014,Seo2019,Baba2020}. This led \citet{Baba2020} to show, using N-body/hydrodynamics simulations, that indeed one could determine the age of the bar by studying the star formation history of an NSD and identifying the boost in star formation corresponding to the bar formation. \citet{Sanders2024} used this method to determine the age of the Milky Way's bar to be $\sim 8$ Gyr. Their observationally determined star formation history shows a boost approximately 8 Gyr ago that is remarkably similar to that predicted by the simulations (compare for example Fig.~\ref{fig:SFH} with Fig.~3 of \citealt{Baba2020}).

The star formation history (SFH) of the NSD contains important information about the accretion history onto a galactic nucleus. The intensity of the bar-driven inflow controls the secular growth of the NSD after bar formation. Periods during which the galaxy and the bar are starved of gas, and the properties of the inflowing gas (e.g.\ metallicity), should be imprinted as a fossil record in the star formation history of the NSD.  For example, in a lenticular galaxy the bar has run out of gas at some point during its history, halting star formation in the NSD. Therefore, the age of the youngest stars in the NSD of a lenticular galaxy should tell us when the galaxy ran out of gas. In this context, it is important to emphasize that not all the inflowing gas is necessarily converted into stars, as a significant portion may be expelled in galactic outflows \citep[e.g.][]{Stuber2021,Ponti2021}.

In the inside-out formation scenario, the age and metallicity of the gas as a function of radius is predicted to reflect the radius of the star-forming nuclear ring that is moving outwards as a function of time. The radius of the ring is correlated with the radius of the Inner Lindblad Resonance. Therefore, by studying the age of the stars (i.e.\ the star formation history) as a function of radius, we can learn how the ILR evolved in time, which tells us important information about the evolution of bar parameters such as its mass distribution and pattern speed. For example, a stronger radial age gradient might point to a fast outward-moving ILR and therefore to a stronger slow-down of the bar pattern speed. In this context, chemical evolution models should prove powerful tools that allow us to predict how metallicity and chemical abundances are distributed in radius. Currently, there are only two chemical evolution models of NSDs available. The first is a one-zone model of the Milky Way's NSD by \citet {Grieco2015}. The second is a multi-zone model within the inside-out framework by \citet{Friske2023}. These works have shown that the current chemical composition of NSDs depends on both the accretion and outflow history of the nuclear region. However, the current lack of chemical abundance data (see Sect.~\ref{sec:abundances} and \ref{sec:externalstellarpopulations}) limits the applicability of the models, which are at the moment largely predictive in nature. Promising prospects for obtaining new data are offered by recent and upcoming facilities (see Sect.~\ref{sec:upcoming}).

The total mass of an NSD provides a lower limit on the total amount of angular momentum that the bar has gained from channelling the gas to the centre of the galaxy. This process tends to increase the bar pattern speed, contrasting the bar slow-down due to dynamical friction with the dark matter halo \citep{Weinberg1985,Athanassoula2003,Chiba2023}. However, the change in the bar pattern speed estimated from NSD masses is usually small. For example, the MW NSD has a mass of $M_{\rm NSD} \simeq 10^9\Msun$, a radius of $R_{\rm NSD} \simeq 100\pc$ and a circular velocity of $v_{\rm NSD} \simeq 100\kms$ (Sect.~\ref{sec:MWkinematics}), while the tips of the MW's bar are approximately at a Galactocentric radius of $R_{\rm bar}\simeq 3\kpc$ and the typical circular gas velocity is $v_{\rm c} \sim 200\kms$. Therefore, the bar has gained an angular momentum of $\Delta L \simeq M_{\rm NSD} ( R_{\rm bar} v_{\rm c} - R_{\rm NSD} v_{\rm NSD}) \simeq  M_{\rm NSD} R_{\rm bar} v_{\rm c} \simeq 6 \times 10^{14} \Msun \pc \kms  $ from the gas that formed the NSD. Crudely estimating the total angular momentum of the bar as $L_{\rm bar} = \alpha M_{\rm bar} R_{\rm bar} \Omega_{\rm p} \sim 10^{18} \Msun \pc \kms  $, where $M_{\rm bar}\simeq 1.8 \times 10^{10}\Msun$ is the mass of the bar, $\Omega_{\rm p}$ its pattern speed \citep{Bland-Hawthorn2016}, and $\alpha$ is a numerical factor related to the moment of inertia of the bar that for the sake of this estimate we approximate as that of a uniform rod as $\alpha=1/12$. Thus, we estimate that in the MW the bar has increased its angular momentum by $0.06\%$ to form the NSD. This is a lower limit since the  mass of the NSD only provides a lower limit on the amount of gas channelled to the centre, as there could have been more gas that instead of forming stars has been expelled by outflows.

\subsection{Formation and destruction of nuclear bars} \label{sec:nbarformation}

The formation and evolution of large-scale galactic bars is a well studied topic in galactic dynamics \citep[e.g.][for reviews]{Sellwood1993,Athanassoula2013}. Bars are believed to form from axisymmetric discs via dynamical instabilities, sometimes triggered by tidal interactions with other galaxies \citep[e.g.][]{Lokas2016,Zana2018}. The formation usually involves two phases: an $m=2$ bar instability that forms a barred structure within the disc, followed by a thickening phase during which the bar thickens into a structure that appears boxy or peanut-shaped in its central parts when observed from the side. The most common studied mechanism for bar thickening is the buckling instability, in which the bar bends out of the disc plane to form a vertically thick structure, but other mechanisms for bar thickening are possible as well \citep{Sellwood2020}.

The formation of nuclear bars is less well understood. Important empirical clues include: 1) Nuclear bars are oriented randomly and rotate faster than outer bars (see Sect.~\ref{sec:nbars}). This indicates that the two bars are dynamically decoupled or at most weakly coupled \citep{Pfenniger1990,Friedli1993,Maciejewski2000,Debattista2007}; 2) We know at least one example, NGC 1291, of a nuclear bar that has a boxy-peanut structure, similar to the one of main bars \citep{Mendez-Abreu2019}. This shows that, similarly to main bars, nuclear bars also thicken, possibly (but not necessarily, \citealt{Sellwood2020}) as a result of buckling instability; 3) Nuclear bars are common stellar structures, being present in approximately 20\% of all barred galaxies, with the fraction rising for more massive galaxies (see \citealt{Erwin2024} and Sect.~\ref{sec:nbars}). This indicates that nuclear bars are either long-lived structures or recurrent structures that reform frequently after dissolution; 4) The stellar population in nuclear bars are younger than those of the outer bar \citep{deLorenzo-Caceres2013,deLorenzo-Caceres2019}, but have similarities to those of outer bars in that they show elevated metallicity and low $[\alpha/{\rm Fe}]$ abundances (\citealt{Bittner2021} and Sect.~\ref{sec:nbars}). Overall, all these observational facts point to nuclear bars being ``scaled-down'' versions of outer bars.

All modern theories for the formation of nuclear bars involve bar instabilities. The common idea is that an NSD, possibly containing gas too, undergoes similar instabilities as the ones that form outer bars. Several simulations have successfully managed to reproduce long-lived nuclear bars in this way \citep[e.g.][]{Rautiainen1999,Rautiainen2002,Debattista2007,Saha2013,Du2015,Wozniak2015,Semczuk2024}. Early numerical simulations \citep[e.g.][]{Friedli1993,Combes1994,Friedli1996} tended to suggest that the presence of gas was essential for the formation of a nuclear bar. More recent simulations have shown that a gas component is not necessary, and nuclear bars can be formed in purely N-body simulations \citep[e.g.][]{Rautiainen1999,Rautiainen2002,Debattista2007,Saha2013,Du2015}.

The simulations above can be divided into two broad categories according to whether the outer bar forms first and the nuclear bar later \citep[e.g.][]{Wozniak2015}, or the opposite \citep[e.g.][]{Rautiainen2002,Du2015,Semczuk2024}. In the first case, the large-scale bar transports gas into the centre, an NSD is formed, and when it has grown sufficiently it can decouple from the outer parts, become self-gravitating and bar-unstable, and form a nuclear bar \citep{Wozniak2015}. In the second case, one starts with two axisymmetric discs of different scale length nested inside each other, and since the dynamical time of the central disc is much shorter, a nuclear bar forms first, while the outer bar forms later either by spontaneous bar instability \citep[e.g.][]{Friedli1993,Du2015} or triggered by tidal interactions \citep{Semczuk2024}. It is possible that some nuclear bars are formed through the first route, while others from the second. The first case fits naturally into the context of bar-driven NSD formation scenario. The second case requires a pre-existing inner disc. \citet{deSa-Freitas2025} finds that nuclear bars generally appear to be younger than main bars, favouring the first scenario. \citet{Semczuk2024} have shown using the TNG50 simulations of galaxy formation that the second scenario can happen in a cosmological context, and we speculate that it might provide an explanation for particularly large ``nuclear'' bars that are as large as the smallest
single bars \citep{Erwin2005}. 

Similar to large-scale bars, nuclear bars can dissolve if the central mass concentration, which can be either a black hole, NSC or gas, grows too big \citep[e.g.][]{Du2017,Li2023,Nakatsuno2024}. This has been proposed to regulate the growth of black holes in some galaxies as the disappearance of the nuclear bar might stop the inner bar-driven gas flow \citep{Du2017}. It has also been proposed that an axisymmetric NSD might result from from the dissolution of a nuclear bar \citep{Wozniak2015}.

\subsection{Relation to the in-situ build up of Nuclear Star Clusters} \label{sec:relationNSC}

The formation of NSDs might be related to the in-situ formation of NSCs in at least two ways. First, although in the bar-driven NSD formation scenario most of the gas inflow stalls at the nuclear ring, a tiny fraction might spill out and continue to the NSC. Hydrodynamical simulations suggest that stellar feedback (e.g.\ supernovae) and/or magnetohydrodynamic turbulence can drive nuclear inflows from the gaseous nuclear ring to the innermost few pc \citep{Tress2020,Moon2023,Tress2024}. This nuclear inflow can create circum-NSDs (CNDs) of gas in the innermost few pc. A particular clear example is provided by NGC3351, which contains a small circum-nuclear ring of radius $R\sim10\pc$ inside the main star-forming nuclear ring at $R\sim 300\pc$, the two being radially separated by a lower-density region (see Fig.~2 of \citealt{Sun2024}). The Milky Way hosts a similar circum-NSD with radius $R\sim4\pc$ \citep[e.g.][and references therein]{Henshaw2023}, which is suggestively similar to the size of the  NSC ($\sim 5\pc$, e.g. \citealt{Gallego-Cano2020}). In-situ star formation in CNDs can grow NSCs, and point to a possible co-evolution of NSCs and NSDs.

The second way in which NSDs and NSCs might be connected is within the context of the inside-out formation scenario (Sect.~\ref{sec:insideout}). If the initial central mass concentration at the moment of bar formation is very small, the radius of the inner Lindblad resonance (ILR) will be very small or non-existent. In this case, the radius of the nuclear ring will be essentially zero, and the bar-driven inflow will deposit the gas at or near the very centre. Only when sufficient mass has accumulated the nuclear ring moves out. In this way, the NSD starts forming from the very centre. This suggests that the NSC and NSD are essentially the same structure, with the NSC simply being the inner part of the NSD \citep{Nogueras2023}.

Putting the two above together, we can construct a possible scenario for the formation of (some) NSCs: the bulk of stars of the NSC forms in the early stages of the inside-out growth. Then in-situ star formation during the following Gyr continues to grow the NSC in co-evolution with the growing edge of the NSD (in this second stage, the NSD and NSC could grow at different rates, see also Sect.\ref{sec:scaling}). This scenario predicts the NSC to be older or equal to the NSD. Measurements of the SFH of NSC and NSD are still too uncertain to confirm or disprove this scenario \citep[e.g]{Nogueras2020NatAs,Chen2023}.

\section{Future Perspectives} \label{sec:perspectives}

\subsection{Open Questions}

In this review, we presented observed properties of NSDs, and discussed the current understanding of the theory of the formation and evolution of NSDs, which has developed particularly in the context of bar-driven evolution. However, the study of NSDs is a relatively new field, and there are plenty of open questions. We outline a selection of these questions below.

\begin{enumerate}
    \item How often are NSDs found in barred and truly unbarred, disc galaxies, and how often are they found in elliptical galaxies? It is clear that observations are limited in terms of sample selection and the quality of the data available to assess the presence of a bar, a main galaxy disc, and the NSD itself. For the latter, in particular, accurate answers require, at a minimum, homogenous, physical spatial resolution across the sample studied. Another important aspect to consider is that, rigorously, the presence of an NSD is better ascertained via measurements of the stellar kinematics. Photometric studies, while an excellent first approximation in this context, are known for systematic effects that are difficult to overcome. A related question is how is the fraction of galaxies with NSDs distributed in terms of galaxy mass? While we have shown above some important initial steps in answering these questions, it is clear that more work is needed to ensure accurate answers. These answers will provide quantitative constraints for models of the formation and evolution of NSDs in galaxies with different structural properties, and may be helpful in understanding the evolution of bars too.

    \item Do NSDs in barred and unbarred galaxies form through the same processes? How do NSDs form in elliptical galaxies? High quality and abundant data is required to understand if the properties of NSDs are different in these different kinds of galaxies. One could envisage a scenario in which NSDs in unbarred disc galaxies have formed through bar-driven evolution just in the same way as in barred galaxies, except that the bar has dissolved by secular evolution or disappeared due to a merger; but is this a plausible scenario? Could the NSD survive the merging process? External gas accretion could be the main mechanism to form NSDs in elliptical galaxies. Cosmological simulations would be very helpful in this context. However, they are currently not quite yet at a level in which NSDs can be studied, given in particular the resolution required to study these structures \citep[see][]{Grand2017,Semczuk2024}.

    \item Cosmological simulations able to reveal the formation and evolution of NSDs are critical to our understanding of the effects of the environment and feedback, for example. These simulations could provide predictions on the distributions of stellar ages, metallicities and abundances, as well as kinematics. Furthermore they could also give clues to the relative importance of multiple formation channels (bar-driven vs. accretion), predictions for scaling relations (Sect.~\ref{sec:scaling}) and insights on the connection between NSDs and NSCs. Finally, such numerical work would be important to assess the robustness of NSDs against mergers.

    \item Does the Milky Way NSD host a nuclear bar? And why does the NSD in the Milky seem to be significantly smaller than NSDs in external galaxies with similar mass and bar length?

    \item What is the connection between the formation and evolution processes of NSDs and NSCs? Do they co-evolve? Or do they have separate formation processes?
\end{enumerate}

\subsection{What can we learn from upcoming facilities?} \label{sec:upcoming}

A number of new, powerful facilities are expected to start operating within the next decade. Their superb characteristics, alongside contemporaneous theoretical efforts, will certainly help in answering the questions outlined above, but we should also expect the unexpected. The unprecedented spatial resolution and statistical power these facilities will provide are bound to lead to unanticipated discoveries.

The Multi-Object Optical and Near-IR Spectrograph (MOONS, \citealt{Cirasuolo2020}) is the forthcoming third-generation instrument to be installed  in 2025 at ESO's VLT in Chile. MOONS will have a spectral resolving power R between 4,000 to 20,000, excellent multiplex capabilities (allowing for the spectra of a thousand objects to be registered simultaneously), and near-IR wavelength coverage. Thanks to these specifications, MOONS at the VLT will be a unique facility for measuring accurate radial velocities, metallicities and chemical abundances for several million stars across the Milky Way. This makes MOONS the ideal instrument to observe stars in the highly-obscured regions of the inner Galaxy and in the Milky Way NSD. A dedicated GTO survey (with 100 nights devoted to science on the  Milky Way, \citealt{Gonzalez2020}) will cover the NSD, besides the Galactic bar and disc, allowing to do for the first time a chemically homogeneous survey by using red clump stars as the main tracer. MOONS will give us more insights on the connections between the NSC, the NSD and the Galactic bar. 

The 4-metre Multi-Object Spectroscopic Telescope \citep[4MOST;][]{deJong2019}, scheduled to start operations in 2026 at ESO's Visible and Infrared Survey Telescope for Astronomy (VISTA) in Chile, will provide an optical counterpart to MOONS. Synergies between the two instruments can be explored in the context of orbital and stellar population substructures particularly in the Galactic bar.

JWST, with its high angular resolution, high sensitivity, its unique wavelength coverage and extremely stable PSF of NIRCAM, has the potential to give us an unprecedented detailed view of the Milky Way's NSD. The proposed JWST Galactic centre survey \citep{SchoedelJWST} aims to measure the physical and kinematics properties of $\sim 10$ million stars in the MW's NSD. If approved, this survey would allow us to answer important questions about the structure and kinematics of the MW's NSD, such as the possible presence of a nuclear bar, to measure in detail its star formation history, and study its connection to the NSC and the overall history of the MW.

The Japan Astrometry Satellite Mission for INfrared Exploration (JASMINE, \citealt{JASMINE}) will be the first dedicated infrared astrometry space mission which will provide high precision astrometry in the Galactic Center allowing to obtain precise proper motions.

In what concerns external galaxies, an ongoing effort from the PHANGS collaboration \citep{Williams2024} is studying NSCs using a combination of several telescopes, including JWST and HST \citep[e.g.][for NGC\,628]{Hoyer2023}.

Given the impact of the MUSE IFS on the study of NSDs in nearby galaxies, it is exciting to note the similar upcoming facilities in store for the VLT and the Extremely Large Telescope \citep[ELT;][]{Gilmozzi2007}. At the VLT, BlueMUSE, an instrument largely based on MUSE but crucially extending the wavelength range down to 350 nm, with a higher spectral resolution, will be installed in the early 2030's  \citep{BlueMUSE}. With these specifications, BlueMUSE will allow chemo-kinematical maps in NSDs of nearby galaxies with improved accuracy, particularly for the younger stellar populations that often reside in the outer edges of NSDs, connecting with the bar. The Multi-conjugate Adaptive optics Visible Imager and Spectrograph (MAVIS, \citealt{MAVIS}) will be installed at the VLT around 2030, using $6-8$ laser guide stars to obtain diffraction-limited images over a $30''$ field-of-view and allowing integral field spectroscopy over a $3''$ field of view. MAVIS will allow to get precise dynamical mass estimates and extend the study of NSCs and NSDs in low-mass galaxies, which is so far lacking.

Looking further ahead in time, the ELT will be equipped in the 2030's with the High Angular Resolution Monolithic Optical and Near-infrared Integral field spectrograph (HARMONI). With high sensitivity and high spatial resolution, HARMONI will allow the study of NSDs in a wide range of sizes and galaxy masses, and in a much larger volume than what is possible today. Furthermore, we will be able to verify the presence of NSDs at cosmic noon ($z\sim1-3$) and test our current understanding of the early formation of bars and the bar-driven build up of NSDs. In fact, HARMONI will allow a TIMER-like study on NSDs at high redshifts. HARMONI will also elucidate the possible connection between the formation of NSDs and NSCs in nearby galaxies. In addition, two other future spectrographs are being planned to be installed in dedicated 12-m class telescopes: the Wide Spectroscopic Telescope \citep[WST;][]{Mainieri2024,Bacon2024} and the Mauna Kea Spectroscopic Explorer \citep[MSE;][]{TheMSEScienceTeam2019}. While the plans for the latter do not include an integral field spectrograph, it is expected to cover the entire optical window from 360–950nm, as well as the J and H near-infrared bands. In addition, it will have over 3000 fibres feeding spectrographs with resolution R ranging from 3000 to 40000. It will thus naturally complement efforts such as those from using 4MOST and/or MOONS on studies of the Milky Way and satellite galaxies. On the other hand, the WST will combine a multi object spectrograph (MOS) model with an integral field unit, both operating in the range from 350nm to 1.6$\mu$m. The MOS is expected to have 20000 fibres providing low-resolution (R $\sim4000$) spectra and 2000 fibres for high resolution spectra (R $\sim40000$). The integral field unit is expected to have a field of view of 3$\times$3 arcmin$^2$ and provide spectra with resolution R $\sim3500$. Therefore, WST will not only provide further insights on the Milky Way central region via the MOS setup, but crucially, the near-infrared capability of the IFS creates the opportunity to understand more accurately the dusty, star-forming NSDs at the centre of spiral galaxies.

\section*{Acknowledgments}
We thank the two anonymous reviewers for their insightful comments, which helped improving the presentation of this review and the discussion of the overall state-of-the-art. We also would like to thank E.M.~Corsini, P.~di Matteo, L.~Morelli, F.~Nogueras-Lara, A.~Pizzella, N.~Ryde,  C.~de S\'a Freitas,  J.L.~Sanders, and   R.~Sch\"odel,  for their feedback and very valuable comments. We are very grateful to F.~Matteucci for her trust and support  to us doing this review paper. MCS acknowledges financial support from the European Research Council under the ERC Starting Grant ``GalFlow'' (grant 101116226) and from Fondazione Cariplo under the grant ERC attrattivit\`{a} n. 2023-3014. DAG acknowledges financial support from STFC grant ST/X001075/1.



\bibliography{sn-bibliography}

\end{document}